\documentclass[aps,preprint,onecolumn,nofootinbib]{revtex4}
\usepackage{tipa,mathrsfs}
\usepackage{bm}% bold math
\usepackage{mathrsfs}
\usepackage{amssymb}
\usepackage{amsmath}
\usepackage{amsfonts,dsfont}
\usepackage{color}
\usepackage{ulem}

\usepackage{graphicx}
\usepackage{amssymb}
\usepackage{amsmath}
\usepackage{amsfonts}
\usepackage{mathrsfs}
\usepackage{color}
\usepackage{ulem}
\usepackage{physics}
\usepackage[english]{babel}
\usepackage[utf8]{inputenc}
\usepackage[T1]{fontenc}
\usepackage{mathptmx}
\usepackage{etoolbox}
\usepackage[]{graphicx}
\begin{document}

\newcommand{\half}{\frac{1}{2}}
\newcommand{\pcl}[1]{#1_{\mathrm{p}}}
\title{Can Bohmian mechanics be considered complete?}
\author{Aur\'elien Drezet $^{1}$, Arnaud Amblard $^{1}$}
\address{$^{1}$\quad Institut N\'eel, UPR 2940, CNRS-Universit\'e Grenoble Alpes, 25 avenue des Martyrs, 38000 Grenoble, France
}

\email{aurelien.drezet@neel.cnrs.fr}
\begin{abstract}
In this work celebrating the centenary of quantum mechanics, we review the principles of de Broglie Bohm theory, also known as pilot-wave theory and Bohmian mechanics.   We assess the most common reading of it (the Nomological interpretation based on the notion of primitive ontology in tridimensional space) and defend instead a more causal and pluralistic approach, drawing on classical analogies with optics and hydrodynamics.   Within this framework, we review some of the approaches exploiting mechanical analogies to overcome the limitations of current Bohmian theory and perhaps quantum mechanics itself.
 
\end{abstract}

\maketitle
%%%%%%%%%%%%%%%
 \section{Introduction}
\indent One hundred years ago, quantum mechanics was founded following Heisenberg's fundamental work on matrix mechanics \cite{Heisenberg,Waerden}. These brilliant results were amplified and developed by Heisenberg in collaboration with Born, Jordan, Pauli, Dirac \cite{Waerden} and under the patronage of Bohr, who in 1927 concluded this prolific period with the introduction of the principle of complementarity \cite{Bohr}, the keystone of what later became known as the Copenhagen school. At the same time, Schr\"{o}dinger, building on de Broglie's ideas, founded wave mechanics in 1926 and introduced the famous equation that bears his name \cite{Schrodinger}. The two approaches of Heisenberg on the one hand and Schr\"{o}dinger on the other differed fundamentally in their methods and objectives.   Whereas Heisenberg, inspired by Mach's positivism and the work of Bohr, Kramers and Sommerfeld, sought to eliminate unobservable quantities from quantum theory, Schr\"{o}dinger wanted to obtain a mechanical image of the electron in the classical sense, but representing it as a vibration extended in space. Schr\"odinger's interpretation of his wave function $\Psi(q,t)$ (defined at time $t$ at coordinate point $q:= [q_1,...,q_{3N}]$ in the $3N$-dimensional configuration space for $N$ particles) was strongly criticized by Heisenberg and Bohr, and the story goes that it was finally Born who officially proposed in 1926 \cite{Born} to interpret the density $|\Psi(q,t)|^2$ as a probability density in configuration space (i.e., eliminating trajectories and determinism from quantum theory).\\
\indent The official, orthodox interpretation of quantum mechanics thus completely absorbed the Schr\"{o}dinger wave equation into its general formalism, providing it with a powerful tool without changing the corpus of principles of the Copenhagen interpretation. 
In fact, in this theory, the probability $|\Psi(q,t)|^2d^{3N}q$ represents the probability of observing at time $t$ the system in the $3N$-dimensional $d^{3N}q$ configuration space element. However, this is by no means a probability of occurrence or presence in the classical sense, as the particles are not assumed to have trajectories or even pre-exist experimental measurement (in this respect, the situation is very different from the classical statistical mechanics used to justify thermodynamics).\\
\indent We could in principle stop our little summary of the great history of quantum mechanics here, but that would of course be a gross oversimplification. Indeed, up to now we've completely overlooked de Broglie, who in 1923 developed a very different image of quantum theory \cite{debroglie1925}.   In fact, unlike the Copenhagen school, which was strongly dominated by the positivist and instrumentalist currents of his time, de Broglie sought to unify quantum theory and classical physics in a mechanical vision in which particles have trajectories but are guided by waves \cite{debroglie1927,Solvay,debroglie1930}.  This essentially deterministic vision differed greatly from Born's probabilistic and stochastic version, which profoundly rejected any return to an approach involving particle trajectories in space-time or configuration space. De Broglie's point of view also differed sharply from Schr\"odinger's approach, which eliminated trajectories and retained only the wave. 

To celebrate the birth of quantum mechanics, we propose to look back at the birth of de Broglie's theory, which has often been overlooked. The aim will be to recall the assumptions on which this original dynamical approach is based, and also to discuss the particular conditions and constraints of this theory that would allow it to be generalized.  Indeed, as de Broglie's approach is essentially based on an analogy with classical mechanics (as we shall recall, starting with the Hamilton Jacobi equation), the question immediately arises as to whether this analogy is complete. Indeed, the Hamilton Jacobi approach, which is linked to fluid dynamics and the Euler-Bernoulli equation, is not the most generic description of an inviscid continuous fluid. In fact, extensions were discussed in the 1950s by Takabayasi \cite{Takabayasi} and Sch\"onberg \cite{Schonberg} after Bohm's rediscovery of de Broglie's theory in 1952 \cite{BohmI}. In this article, we would like to re-analyze these somewhat forgotten approaches in the context of modern de Broglie-Bohm theory \cite{BohmHiley,Holland,Teufel,Durr}.

 This article is organized as follows: In section \ref{section2}, we will review the basics of de Broglie Bohm theory, with particular reference to Madelung's hydrodynamic formalism. In section \ref{section3}, we discuss the 'mainstream' interpretation of de Broglie Bohm's theory at present, i.e. the so-called 'nomological' interpretation of D\"urr, Goldstein Zangh\`i (DGZ) \cite{DGZ1,Allori}.  We criticize this interpretation and show its physical limitations.  In section \ref{section4}, we show how we can move beyond the nomological interpretation and seek to extend the accepted Bohmian theory. In this section, we focus on the Bohm-Vigier research program  \cite{Bohmchaos,Vigier} taken up by Valentini \cite{Valentini,Valentini2020}, which seeks to relax the statistical equilibrium conditions that usually give the famous Born rule in $|\Psi|^2$.
In section \ref{section5}, we look at interpretations that seek to modify Bohm's theory from a dynamic point of view. This essentially includes Bohm's approaches involving stochastic terms.   However, we are rather critical of this program and prefer to consider in section \ref{section6} the extension of de Broglie Bohm theory involving a vorticity field associated with particle velocity. We show that this type of theory could nicely extend the framework of the de Broglie-Bohm approach, while defining a field that could potentially link up with the research proposed by Bohm-Vigier and Valentini. Finally, we conclude and summarize our analysis in section \ref{section7}.  
%%%%%%%%%%   
\section{The de Broglie-Kennard-Bohm theory and the Madelung hydrodynamical approach} \label{section2}
\indent First, we will briefly summarize the method de Broglie developed to obtain his wave-guided particle mechanics, based on Schr\"{o}dinger's wave theory. In fact, both de Broglie's \cite{Solvay,debroglie1930} and Schr\"{o}dinger's \cite{Schrodinger} approach made extensive use of classical analogies with the famous Hamilton-Jacobi (HJ) mechanics, where classical particles are guided by a function $S(q,t)$ known as the HJ action (for a technical discussion see \cite{Landau,GoldsteinB,Brillouin,Holland}). In this formalism, the guiding formula for the system described by the position variable $q(t)$ is given by:
\begin{eqnarray}
m_k\frac{d}{dt}q_k(t)=\partial_kS(q(t),t)\label{1}
\end{eqnarray}
with $\partial_k:=\frac{\partial}{\partial q_k}$ and $m_k$ the particles masses $k=1,..., 3N$ (here $m_k$ are degenerated three by three: $m_1=m_2=m_3$,...,$m_{3N-2}=m_{3N-1}=m_{3N}$).
This dynamics corresponds to the nonrelativistic HJ equation 
\begin{eqnarray}
-\partial_tS(q,t)=\sum_k \frac{1}{2m_k}(\partial_kS(q,t))^2+V(q,t)\nonumber\\
=\sum_k\frac{m_k}{2}(\dot{q}_k)^2+V(q,t)
\label{2}
\end{eqnarray} 
in the presence of external potential $V(q,t)$ (we use $\partial_t:=\frac{\partial}{\partial t}$). $-\partial_t S $ plays the role of the energy and the HJ relation is more generally written  $-\partial_t S =H(q,p,t)$ with $H$ the Hamiltonian and $p=\partial_q S (q,t):=[p_1=\partial_kS(q,t),...,p_{3N}=\partial_{3N}S(q,t)]$ the momenta of the particles. \\
\indent De Broglie's beautiful central idea is to retain the guiding formula Eq.~\ref{1} but modify the HJ wave equation Eq.~\ref{2} by replacing it with the Schr\"{o}dinger equation
\begin{eqnarray}
i\partial_t\Psi(q,t)=\sum_k \frac{-1}{2m_k}\partial^2_k \Psi(q,t)+V(q,t)\Psi(q,t)\label{3}
\end{eqnarray}
(in the following we assume natural units $\hbar=c=1$). The mathematical development begins with the polar equation $\Psi(q,t)=R(q,t)e^{iS(q,t)}$ ($R$ and $S$ being real functions of $q$ and $t$), which is inserted into Eq.~\ref{3} to obtain the pair of equations:
\begin{eqnarray}
-\partial_tS(q,t)=\sum_k \frac{1}{2m_k}(\partial_kS(q,t))^2+V(q,t)+V_\Psi(q,t)\label{4}\\
\sum_k\partial_k(R^2(q,t)\frac{\partial_k S(q,t)}{m_k})+\partial_tR^2(q,t)=0.\label{5}
\end{eqnarray}
Clearly Eq.~\ref{4} is riminiscent of Eq.~\ref{3} with the additional quantum potential
\begin{eqnarray}
V_\Psi(q,t)=\sum_k \frac{-1}{2m_k}\frac{\partial^2_k R(q,t)}{R(q,t)}\label{6}
\end{eqnarray} which disappears if the Planck constant $\hbar$ vanishes. The dynamics proposed by de Broglie is thus given by a quantum version of the HJ equation involving $V_\Psi(q,t)$. The guidance formula Eq.~\ref{1} suggests itself where $S(q,t)$ is now the phase of the wave function $\Psi$ (i.e., $S=-\frac{i}{2}\log{(\Psi/\Psi^\ast)}$)
\begin{eqnarray}
p_k=m_k\frac{d}{dt}q_k(t)=\partial_kS(q(t),t)=\textrm{Im}[\frac{\partial_k\Psi(q,t)}{\Psi(q,t)}]\label{7}.
\end{eqnarray}  
De Broglie's particle trajectories are obtained by integration of the first order differential equations 
\begin{eqnarray}
\frac{m_1dq_1}{\partial_1S(q(t),t)}=...=\frac{m_{3N}dq_{3N}}{\partial_{3N}S(q(t),t)}=dt.
\end{eqnarray} 
An important relation is obtained by taking the gradient $\partial_j$ of Eq.~\ref{4}: 
\begin{eqnarray}
\frac{d}{dt}v_j(q,t)=(\partial_t+\sum_kv_k(q,t)\partial_j)v_j(q,t)=-\frac{1}{m_k}\partial_j(V(q,t)+V_\Psi(q,t))\label{Newton}
\end{eqnarray} 
This is Newton's (second order) law of motion for the particles in presence of $V(q,t)$ and $V_\Psi(q,t)$. Naturally in absence of quantum potential we recover classical mechanics which is just another way of saying that classical (or quantum) HJ formalism agrees with Newtonian formalism in terms of force and acceleration. It should be noted that the fact that Newton's laws can be recovered from the HJ formalism does not imply that the two approaches, HJ and Newton, are equivalent. Indeed, the HJ formalism only considers a restricted class of motions and trajectory distributions in configuration space that are compatible with Newton's laws (we will come back to this issue later). 

An other remarkable property of de Broglie's approach is that it gives immediate meaning to Eq.~\ref{5}, which can be written: 
\begin{eqnarray}
\sum_k\partial_k(R^2(q,t)v_k(q,t))+\partial_tR^2(q,t)=0.\label{8}.
\end{eqnarray} where $v_k(q,t)$ is an Eulerian velocity field for the probability fluid with density $\rho_\Psi(q,t):=|\Psi(q,t)|^2$. The Eulerian velocity is in fluid dynamics equals to the Lagrange particle velocity $\frac{d}{dt}q_k(t):=\dot{q_k}(t)$ along 	a given trajectory of the system (i.e., $\frac{d}{dt}q_k(t)=v_k(q(t),t)$). Therefore the velocity of the probablity fluid coincids with the de Broglie guidance formula obtained from the HJ quantum equation Eq.~\ref{4}. \\
\indent  From a historical point of view, it's worth noting that although de Broglie began working on his approach as early as his thesis work in 1923-1925 \cite{debroglie1925}, he didn't introduce the polar representation until 1925-26 in the context of its double solution theory \cite{debroglie1926a,debroglie1926b,debroglie1927} (we will go back to this point later) and only made full use of it when Schr\"{o}dinger obtained his equation. In 1926 Brillouin  \cite{brillouin1926}(who worked in the same laboratory as de Broglie) used polar notation to develop the famous WKB (Wentzel-Brillouin-Kramers) semiclassical approximation, starting from a $\hbar$ power expansion of quantum HJ.  The exact representation given by Eqs. \ref{4} and \ref{5} was not introduced by de Broglie until 1927 \cite{debroglie1927}, simultaneously with Madelung who started working on it in 1926 \cite{Madelunga,Madelungb} (and the representation is also called Madelung representation for this reason). De Broglie applied it to both the non-relativistic Schr\"odinger equation (for $N$ particles) \cite{debroglie1927,Solvay,debroglie1930} and the relativistic version based on the Klein-Gordon equation (which was actually first published by de Broglie) for one particle \cite{debroglie1927b,debroglie1927c}.\\
\indent De Broglie presented his theory at the famous Solvay Congress in 1927 \cite{Solvay}, and it was discussed by Lorentz, Kramers, Pauli, Ehrenfest, Brillouin and Einstein. Despite Einstein's and Brillouin's support in principle, the community of physicists working around Bohr and Born (which included the entire Copenhagen school) as a whole rejected the value of an approach based on the notion of a deterministic trajectory. Moreover, the non-intuitive nature of Broglian dynamics did not militate in its favor. As a result, de Broglie  abandoned his project in 1928.\\ 
\indent Remarkably, de Broglie's pilot-wave theory was nevertheless rediscovered several times in the course of the twentieth century.  
The most famous of these rediscoveries was of course Bohm's in 1951-52 \cite{BohmI,Bohm1951}, and for this reason the theory is called de Broglie-Bohm.   However, before Bohm, it was also rediscovered by Kennard in 1928 \cite{Kennard} and critically analyzed by Rosen in 1945 \cite{Rosen}. In particular, Kennard, before Bohm \cite{BohmII}, proposed the first application of the theory to explain quantum measurement processes \cite{Kennard}. The important point is that both Kennard and Rosen do not refer to de Broglie, but cite Madelung's work. So it's worth saying a word here about Madelung's hydrodynamical theory, which is formally (but not physically) equivalent to de Broglie's in the non-relativistic domain.\\ 
\indent   In fact, Madelung's theory \cite{Madelunga,Madelungb} differs essentially only in the physical interpretation of trajectories and dynamics calculated from the guidance formula and the HJ equation. More precisely, Madelung adopts the purely wave Schr\"{o}dinger interpretation for a delocalized electron, and interprets the $\rho(q,t)$ distribution not as a probability density, but rather $e\rho(q,t)$ as the charge density of a hypothetical electron fluid ($e$ being the full electron charge).  This perspective is now completely outdated but the hydrodynamical formalism is still valuable in Bohmian mechanics.   The description makes use of the fact that formally the HJ equation is equivalent to the generalized Euler-Bernoulli equation for an inviscid fluid without vorticity. 
To motivate this result we remind \cite{Batchelor} that in classical hydrodynamics (in 3D), an Eulerian inviscid fluid obeys a pair of equations:
\begin{eqnarray}
\frac{d}{dt}\mathbf{v}(\mathbf{r},t)=(\partial_t+\mathbf{v}(\mathbf{r},t)\cdot\boldsymbol{\nabla})\mathbf{v}(\mathbf{r},t)=-\boldsymbol{\nabla}\frac{V(\mathbf{r},t)}{m}-\frac{1}{\mu(\mathbf{r},t)}\boldsymbol{\nabla}P(\mathbf{r},t)\label{9}\\
\partial_t\mu(\mathbf{r},t)+\boldsymbol{\nabla}\cdot(\mu(\mathbf{r},t)\mathbf{v}(\mathbf{r},t))=0\label{10}
\end{eqnarray}
with  $\bm{\nabla}:=\frac{\partial}{\partial\mathbf{r}}$, and where $\mathbf{v}(\mathbf{r},t)$, $\mu(\mathbf{r},t)=m\rho(\mathbf{r},t)$, $V(\mathbf{r},t)$, $P(\mathbf{r},t)$ are respectively the velocity field, the mass density (each particle having the mass $m$), the external potential,  and the pressure field defined at position $\mathbf{r}=[x,y,z]$ and time $t$ in the fluid.
Eq. \ref{9} is of course Euler's equation (i.e. Newton's dynamical law for the local fluid) and Eq.~\ref{10} is a local mass conservation law. Furthermore we now assume a barotropic fluid, $P=P(\rho)$, and we have $\frac{1}{\rho}\boldsymbol{\nabla}P=\boldsymbol{\nabla}F$ with the function $F(\rho)=\int^\rho dp/\rho$. After using the identity $\boldsymbol{\nabla}(\frac{v^2}{2})=(\mathbf{v}\cdot\boldsymbol{\nabla})\mathbf{v}+\mathbf{v}\times\boldsymbol{\Omega}$  (where $\boldsymbol{\Omega}=\boldsymbol{\nabla}\times\mathbf{v}$ is the local vorticity) we deduce:
\begin{eqnarray}
\partial_t\mathbf{v}+\boldsymbol{\nabla}[\frac{v^2}{2}+\frac{F+V}{m}]=\mathbf{v}\times\boldsymbol{\Omega}\label{11}
\end{eqnarray}
and therefore, taking the curl of Eq.~\ref{11}, we obtain the evolution equation for the vorticity field 
\begin{eqnarray}
\partial_t\boldsymbol{\Omega}=\boldsymbol{\nabla}\times(\mathbf{v}\times\boldsymbol{\Omega})\label{12}\\
=(\boldsymbol{\Omega}\cdot\boldsymbol{\nabla})\mathbf{v}-\boldsymbol{\Omega}(\boldsymbol{\nabla}\cdot\mathbf{v})-(\mathbf{v}\cdot\boldsymbol{\nabla})\boldsymbol{\Omega}\label{13}
\end{eqnarray}
Clearly, from Eq.~\ref{12} we see that if the vorticity field is vanishing, $\boldsymbol{\Omega}=0$, at a given time this will be so at any other time.  
A stronger result is obtained with  Eq.~\ref{13} rewritten as $\frac{d}{dt}\boldsymbol{\Omega}=(\boldsymbol{\Omega}\cdot\boldsymbol{\nabla})\mathbf{v}-\boldsymbol{\Omega}(\boldsymbol{\nabla}\cdot\mathbf{v})$ or with Eq.~\ref{10}  $\frac{d}{dt}(\frac{\boldsymbol{\Omega}}{\rho})=(\frac{\boldsymbol{\Omega}}{\rho}\cdot\boldsymbol{\nabla})\mathbf{v}$ from which we deduce that once the condition $\boldsymbol{\Omega}(\mathbf{r},t)=0$ is assumed at a given space-time point $[\mathbf{r}:=\mathbf{R}(t),t]$ then  $\boldsymbol{\Omega}(\mathbf{r'},t')=0$ still holds true	at any other space-time point $[\mathbf{r}':=\mathbf{R}(t'),t']$ belonging to the Lagrangian trajectory of the fluid particle $\mathbf{R}(t)$.  Considering an irrotational fluid, we can thus write 
\begin{eqnarray}
\mathbf{v}(\mathbf{r},t)=\boldsymbol{\nabla}\frac{S(\mathbf{r},t)}{m}\label{14}
\end{eqnarray}
where $S(\mathbf{r},t)/m$ plays the role of velocity potential. Eq.~\ref{14} is formally identical to the guidance formula used in the HJ formalism. Moreover, Eq.~\ref{11} can be rewritten $\boldsymbol{\nabla}[\partial_tS+\frac{mv^2}{2}+F+V]=0$ from which we get  the generalized Euler-Bernoulli formula: 
\begin{eqnarray}
-\partial_tS(\mathbf{r},t)=\frac{(\boldsymbol{\nabla}S(\mathbf{r},t))^2}{2m}+F(\mathbf{r},t)+V(\mathbf{r},t)+f(t)\label{15}
\end{eqnarray} where $f(t)$ is an arbitrary function of time which can be absorbed in the definition of $S(\mathbf{r},t)$. When this is done Eq.~\ref{15} is formally identical to the classical HJ Eq.~\ref{3}    (in 3 dimensions) up to the term $F(\mathbf{r},t)$ associated with the `internal energy' of  the fluid. Note that we can be more explicit concerning the physical meaning of the function $F$. More precisely in the steady regime where the various fields $P,\rho, \mathbf{v}$ are not explicit functions of time $t$ we can derive from thermodynamics the relation $\boldsymbol{\nabla}(\frac{h}{\rho})=\frac{1}{\rho}\boldsymbol{\nabla}P+T\boldsymbol{\nabla}(\frac{\eta}{\rho})$ where $\eta(\mathbf{r})$ is the density of entropy (entropy per unit volume) and $h(\mathbf{r})=\epsilon(\mathbf{r}) +P(\mathbf{r})$ is the local density of enthalpy sum of the density of internal energy $\epsilon$ and local pressure $P$. In the adiabatic case with $\eta/\rho$ constant we have therefore $\boldsymbol{\nabla}(\frac{h}{\rho})=\frac{1}{\rho}\boldsymbol{\nabla}P$. In this steady regime we have $\frac{h}{\rho}:= F=\int^\rho dp/\rho$ which shows that $F$ is actually a kind of internal enthalpy. \\
\indent Madelung's formalism builds on this hydrodynamic analogy and introduces a quantum fluid of density $\rho(\mathbf{r},t)=R^2(\mathbf{r},t)$ associated with Schr\"odinger's delocalized electron but replaces the isotropic pressure field $P(\mathbf{r},t)$ of standard hydrodynamics with a stress tensor $\sigma_{ij}(\mathbf{r},t)$ such that $\boldsymbol{\nabla}P(\mathbf{r},t)$ is replaced by $\boldsymbol{\nabla}\cdot \stackrel{\leftrightarrow}{\sigma}(\mathbf{r},t)$  in Eq.~\ref{12}. In order to recover Eq.~\ref{Newton} in 3D we must have $\frac{1}{\rho}\boldsymbol{\nabla}\cdot \stackrel{\leftrightarrow}{\sigma}=\boldsymbol{\nabla}V_\Psi$ with $V_\psi=\frac{-1}{2m}\frac{\boldsymbol{\nabla}^2R}{R}$. The simplest choice is:
\begin{eqnarray}
\sigma_{ij}(\mathbf{r},t)=\frac{-\rho(\mathbf{r},t)}{4m}\partial_{i}\partial_j\log{(\rho(\mathbf{r},t))}\label{17}
\end{eqnarray}      
Using Eq.~\ref{17} the pair of hydrodynamical relations \ref{9}, \ref{10} becomes:
\begin{eqnarray}
\frac{d}{dt}\mathbf{v}(\mathbf{r},t)=-\boldsymbol{\nabla}\frac{V(\mathbf{r},t)}{m}-\frac{1}{\mu(\mathbf{r},t)}\boldsymbol{\nabla}\cdot \stackrel{\leftrightarrow}{\sigma}(\mathbf{r},t)=-\boldsymbol{\nabla}\frac{V(\mathbf{r},t)+V_\Psi(\mathbf{r},t)}{m}\label{18}\\
\partial_t R^2(\mathbf{r},t)+\boldsymbol{\nabla}\cdot(R^2(\mathbf{r},t)\mathbf{v}(\mathbf{r},t))=0\label{19}
\end{eqnarray}
The analogy with standard and phenomenological hydrodynamics is very strong with however  the function $F$ replaced by the quantum potential $V_\Psi$. Therefore Eqs.~\ref{11}-\ref{13} still hold. Madelung effectively considered an irrotational fluid, $\boldsymbol{\Omega}=\boldsymbol{\nabla}\times\mathbf{v}=0$ which implies a velocity potential $S(\mathbf{r},t)/m$ such that the guidance formula \ref{14} and Euler-Bernoulli law  Eq.~\ref{15} hold true.\\
\indent Therefore having obtained the pair of equations
\begin{eqnarray}
-\partial_tS(\mathbf{r},t)=\frac{(\boldsymbol{\nabla}S(\mathbf{r},t))^2}{2m}+V_\Psi(\mathbf{r},t)+V(\mathbf{r},t)\label{20}\\
\partial_t R^2(\mathbf{r},t)+\boldsymbol{\nabla}\cdot(R^2(\mathbf{r},t)\boldsymbol{\nabla}\frac{S(\mathbf{r},t)}{m})=0
\end{eqnarray} and used the definition $\Psi=\sqrt{\rho}e^{iS}$ we have, with Madelung, recovered Schr\"odinger's equation for a single electron and built a hydrodynamical picture of $\Psi$.\\
\indent Note that Madelung's hydrodynamic formalism can be generalized to the $N$-electron problem (in $3N$-dimensional configuration space), although Madelung himself, inspired by the work of Schr\"odinger, who in 1926-27 hoped to get rid of configuration space in fine, did not do so. Making the link with the de Broglie approach presented earlier, the general idea is to start from the ``hydrodynamic'' equations in configuration space: 
\begin{eqnarray}
\frac{d}{dt}p_k(q(t),t):=(\partial_t+\sum_j \frac{p_j}{m_j}\partial_j)p_k(q,t)=-\partial_k(V(q,t)+V_\Psi(q,t))\label{21}\\
\sum_k\partial_k(R^2(q,t)\frac{p_k}{m_k})+\partial_tR^2(q,t)=0.\label{22}
\end{eqnarray}
with $p_k(q,t)=m_kv_k(q,t):=m_k\frac{d}{dt}q_k(t)$ the particle momenta.
Generalizing Eq.~\ref{11} we obtain the relation 
\begin{eqnarray}
\partial_tp_k+\partial_k(\sum_j \frac{p_j^2}{2m_j}+V+V_\Psi)=-\sum_j \frac{p_j}{m_j}\omega_{jk}
\end{eqnarray}
with the impulse vorticity in the configuration space $\omega_{jk}=\partial_jp_k-\partial_kp_j$.
We deduce the vorticity equation
\begin{eqnarray}
\partial_t\omega_{ik}=-\partial_i(\sum_j \frac{p_j}{m_j}\omega_{jk})+\partial_k(\sum_j \frac{p_j}{m_j}\omega_{ji})
\end{eqnarray}  
which generalizes Eq.~\ref{12}. With Madelung we can postulate an irrotational fluid, i.e. one that cancels out the impulse vorticity:  $\omega_{ij}(q,t)=0$. This allows a gradient field to be created $v_k(q,t)=\frac{p_k}{m_k}=\frac{\partial_k S(q,t)}{m_k}$, in agreement with de Broglie-Bohm guidance formula,  and the Euler-Bernoulli equation equivalent to the HJ equation of the Eqs. pair \ref{4},\ref{5} to be justified by integration.\\
\indent To conclude this section, we'd like to make a few important remarks. Firstly, in our descriptions of de Broglie and Madelung's methods in relation to HJ and Euler equations, we have not sought to be rigorous about the equivalence between the approaches. An interesting point to note, however, is that the transition from classical HJ to Schr\"odinger equation is symmetrical in the sense that if we admit the HJ relations \ref{2} and if we adjoint a probability density $\rho(q,t)$ obeying local conservation $\sum_k\partial_k(\rho(q,t)\frac{p_k}{m_k})+\partial_t\rho(q,t)=0$ in configuration space, then it is possible in classical physics to formally define a complex number $\Psi_{class. }(q,t)=\sqrt{\rho(q,t)}e^{iS(q,t)/a}$ (with $a$ a dimensionless constant, as we have posited $\hbar=1$, if we restore the constant $\hbar$ then $a$ must be replaced by $a\hbar$) satisfying the nonlinear Schrodinger equation:
\begin{eqnarray}
ia\partial_t\Psi_{class.}(q,t)=\sum_k \frac{-a^2}{2m_k}\partial^2_k \Psi_{class.}(q,t)+V(q,t)\Psi_{class.}(q,t)-V_{\Psi_{class.}}\Psi_{class.}(q,t)\label{3New}
\end{eqnarray} wih the classical analog of the quantum potential $V_{\Psi_{class.}}=\sum_k \frac{-a^2}{2m_k}\frac{\partial^2_k \sqrt{\rho}(q,t)}{\sqrt{\rho}(q,t)}=\sum_k \frac{-a^2}{2m_k}\frac{\partial^2_k |\Psi_{class.}|(q,t)}{|\Psi_{class.}|(q,t)}$.
The presence of this potential in Eq.~\ref{3New} is necessary in order to recover the classical HJ relation \ref{2} (this issue was discussed by Schiller \cite{Schiller}, Rosen \cite{Rosenb}, Holland \cite{Holland}, and Vigier \cite{Vigierthesis}).  This shows once more the equivalence between different representations (involving  wave functions or hydrodynamical variables) not only in  quantum but also in classical physics.\\
\indent Another important remark (connected to the previous ones) concerns a postulate which plays a central role in quantum mechanics and which we have so far omitted to discuss. It is indeed  central to all quantum mechanical problems to assume that the wave function $\Psi(q,t)$ is continuous, regular and single-valued. This point is trivially accepted in textbooks and articles, but it implies that in the Madelung de Broglie representation, the phase or action $S(q,t)$ can only be defined to within $2\pi$. Specifically, as noted by Takabayasi \cite{Takabayasib} (see also Holland \cite{Holland}, Berry \cite{Berry} and Bialinicky-Birula \cite{Birulaa,Birulab}), if we integrate the field $p=\nabla S$ along a closed contour $C$ in the configuration space, we must have the quantization condition for the circulation: 
\begin{eqnarray}
\oint_C \sum_k p_kdq_k=2\pi N \label{kelvin1}
\end{eqnarray}
with $N$ an integer, which clearly generalizes the semiclassical Bohr-Sommerfeld formula. A non-zero integer $N$ reveals the presence of phase singularities (or vortices) in the configuration space. These vortices  can only appear at points $q$ where the wave function $\Psi(q,t)=A(q,t)+iB(q,t)$ ($A, B\in \mathbb{R}$) cancels out, which occurs at the intersection of surfaces $A(q,t)=B(q,t)=0$ along open or closed nodal lines where the phase is undefined. In keeping with the non-rotational nature of the Madelung fluid, this implies (by analogy with magnetostatics) a localized current term along singularities. According to Holland \cite{Holland}, in the 3D case we have 
\begin{eqnarray}
m\boldsymbol{\Omega}(\mathbf{x},t)=m\boldsymbol{\nabla}\times\mathbf{v}(\mathbf{x},t)=\sum_a 2\pi N_a \int_{L_a}\frac{\partial\mathbf{z}_a(\lambda_a,t)}{\partial\lambda_a}\delta^{3}(\mathbf{x}-\mathbf{z}_a(\lambda_a,t))d\lambda_a \label{Kelvin}
\end{eqnarray}  where $\mathbf{z}_a(\lambda_a,t)$ are coordinates of a point  on the  $a^{th}$ nodal line $L_a$  ($\lambda_a$ is a parameter) and $N_a$ is an integer characterizing the vortex. Note that following Kelvin's theorem in hydrodynamics the integral  \ref{kelvin1} is a constant of motion: the vortex can change its shape  with time but the integral \ref{kelvin1}  (i.e., the circulation along a loop) will be preserved and carried with the local Madelung flow. This is clearly a topological property of the Madelung fluid. We stress that the need for the condition \ref{Kelvin} in quantum mechanics has sometimes been used by some \cite{Takabayasib,Wallstrom} to argue that Madelung's hydrodynamic formalism and the Schrödinger equation are not equivalent. In our view, however, the controversy only shows that the $\Psi$ condition of continuity and regularity, implying \ref{kelvin1}, \ref{Kelvin} is not imposed by maths but by physics. For example the condition \ref{kelvin1} is central to explain angular momentum quantization in atoms.   Eq. \ref{Kelvin}, is therefore a topological physical property that must be postulated in quantum mechanics to agree with experiments. It is also notable that such a quantization condition is not necessary in classical physics within the framework of HJ dynamics or Eulerian fluid mechanics (the classical function $S(q,t)$ has not in general to be continuous up to $2\pi$). This is an important point, and we'll come back to its significance later.   

%%%%%%%%%%%%%%%%%%%
\section{The nomological interpretation and its problems}
\label{section3}
Following the results obtained in the previous section, we can clearly see the emergence of interesting problems concerning the foundations of the de Broglie Bohm theory.   We have seen with de Broglie that it is possible to define quantum dynamics for material points guided by the phase $S(q,t)$ of the wave function $\Psi(q,t)$. This phase obeys a generalized HJ equation involving a quantum potential $V_\Psi(q,t)$ acting in configuration space and whose expression is highly non-classical. Moreover, to achieve equivalence with the Schr\"odinger equation, we must also impose the conservation relation for the fluid of density $R^2(q,t)$ in configuration space. In this way we have the trio of equations \ref{4}, \ref{5} and \ref{6} defining the de Broglie-Bohm dynamics.\\
\indent However, it is at this point that divergent interpretations arise, and different `Bohmians' or `Broglians' differ as to the right axioms to choose for the theory. \\
\indent One of the most popular interpretations, dating back to the work of de Broglie and Bohm, compares $\Psi$ to a pilot wave carrying a solid particle (like a surfer on his wave or a ``tracer'' following the hydrodynamical flow). Although the analogy is useful for a single particle, it poses problems for the $N$-body case, as the $\Psi(q,t)$ wave generally moves in configuration space inducing nonlocality between particles (making the pilot wave image less appealing).  In particular, the status of hypothetical “empty waves” (and their undetectability) containing no particle (and therefore energy) is still hotly debated \cite{debroglie1956,Selleri,Croca,Hardy,Vaidman,Drezet2006}.\\ 
\indent In this context, the interpretation or reading most often cited today is probably that associated with the seminal work of DGZ \cite{DGZ1992,DGZ1996,GoldsteinZanghi,Goldstein1998,DGZ1,Allori,Esfeld}, who, drawing on the work of John Bell \cite{Bell} (at least at a formal level), have very cleverly sought to define a minimalist formulation of what they call Bohmian mechanics (moreover in this work we often use the expression Bohmian mechanics without relation to the DGZ framework). In this approach, it is no longer necessary to invoke the HJ equation, the conservation of probability fluid constraint or the polar expression of $\Psi$, and it suffices to take as a starting axioms the guiding formula and the Schr\"odinger equation: 
\begin{eqnarray}\left\{\begin{array}{ll}
i\partial_t\Psi(q,t)=\sum_k \frac{-1}{2m_k}\partial^2_k \Psi(q,t)+V(q,t)\Psi(q,t)\\
m_k\frac{d}{dt}q_k(t)=\textrm{Im}[\frac{\partial_k\Psi(q,t)}{\Psi(q,t)}]
   \end{array}\right.\label{26}
   \end{eqnarray}
The great austerity and conciseness of this approach explains its pedagogical and philosophical interest.  Indeed, by eliminating any reference to the HJ or Newton equation at a fundamental level, we obtain a description of de Broglie Bohm trajectories reduced to a simple algorithm free of classical metaphysical prejudice. This greatly simplifies the introduction of this theory at an elementary level. On the other hand, Goldstein Tumulka and  Zangh\`i \cite{GoldsteinTumulkaZanghi} (but also Valentini \cite{Valentini,Valentini2020} and de Broglie \cite{debroglie1956,Vigierthesis}) have criticized the quantum potential $V_\Psi$ for its mysterious nature. It is in fact very different from a traditional force potential such as the Coulomb or gravitational force. Indeed, the  quantum potential $\sum_k \frac{-1}{2m_k}\frac{\partial^2_k R(q,t)}{R(q,t)}$ is expressed as a ratio involving the $|\Psi(q,t)|$ norm, and is unchanged if the wave function is multiplied by an arbitrary constant. What's more, it acts in the configuration space dependent on all positions $q_1...,q_N$ in a highly non-local manner. It has no universal expression (unlike the gravitational or electrostatic potential). It does not weaken with distance in a trivial way, can act in a very specific way between two distant particles but can spare neighboring ones, and does not appear to have a source (unlike, for example, electric or gravitational potential).  The elimination of any reference to $V_\Psi$ and HJ equation therefore seems a good thing, and this argument has been very often taken up by Bohmians and many philosophers.\\
\indent It's clear, however, that the system of equations \ref{26} is disymmetrical, since on the one hand the Schr\"odinger equation is sufficient on its own (by imposing boundary conditions and adding a continuity postulate of $\Psi$) to determine the evolution of the wave function. In contrast, particle motion $q(t)$ is affected by $\Psi$. There is no back-action (reaction) of the particle on the wave, contrary to what we might expect from any mechanical explanation involving the interaction of a $\Psi$ field and particles.  Moreover, particles are of course definable both in configuration space and in three-dimensional physical space. This is not the case for the wave function, which is defined only in configuration space. This seems to introduce a strange ontological gap \cite{Callender}: how can a theory be built with variables acting in different spaces?  It is of course possible to say that the real fundamental space is the configuration space (this is the choice of the philosopher David Albert \cite{Albert}, for example), but this doesn't convince everyone. The configuration space depends on the number of particles presented. Is it realistic to envisage a universe in which the number of dimensions changes as the number of particles increases or decreases? This goes against de Broglie's remark \cite{debroglie1930,debroglie1956}: to have a space of configurations, there must be configurations; in the sense that the space of configurations presupposes the existence of three-dimensional physical space, and not the other way round.  To avoid all these problems, DGZ \cite{DGZ1,GoldsteinZanghi} (see also \cite{Allori,Esfeld}) have proposed interpreting the $\Psi$ wave not as a physical field variable, but as a dynamic or nomological variable, i.e. one associated solely with the notion of a law of motion.  
According to this approach, the `primitive' variables of the theory are the $q(t)$ coordinates of the particles located in space-time. The wave function doesn't have the same status, and is rather comparable to the Hamitonian of classical physics $H(q,p)$ which, via first-order equations $\dot{q}=-\partial H(q,p)/ \partial p$, $\dot{p}=-\partial H(q,p) /\partial q$ determines the motion of particles.\\
\indent One problem with this analogy, however, is that the classical Hamiltonian is determined by physical laws and does not change contingently (it could depend on time, if we introduce external fields but this is also fixed by the laws of physics). This is clearly not the case for the $\Psi(q,t)$ wave function, which is a solution of the Schr\"odinger equation and depends on  initial (and boundary) conditions $\Psi(q,t=0)$. The $\Psi$ variable therefore obeys a dynamic, and this regression of a law dependent on other laws seems hard to swallow.  For this reason, DGZ and the followers of the nomological approach seek to eliminate the contingency associated with the initial wave function and rely in particular on the wave function of the universe $\Psi_{U}$, which is supposed to be a solution of the very hypothetical quantum gravity \cite{GoldsteinTeufel}. If this wavefunction depends, for example, on the Wheeler DeWitt equation, DGZ assumes that this wavefunction is unique, thus eliminating contingency. However, on the one hand, quantum gravity remains speculative and, on the other, it is now accepted since the work of Hartle and Hawking and Vilenkin that the choice of initial or boundary conditions fixing $\Psi_U$ is not unique \cite{Vilenkin}. This strongly weakens the position of DGZ and their collaborators. 

Apart from the reliance on a hypothetical quantum gravity fixing the uniqueness of the universe's wave function, there are other, more serious problems which we believe undermine the nomological interpretation of Bohmian mechanics. DGZ's insistence on an analogy with the classical Hamiltonian $H(q,p,t)$ overlooks the fact that the closest mathematical object to the wave function in classical physics is the action $S(q,t)$ given by the HJ equation \ref{2}.\\
\indent Indeed, comparing the system of equations fixing the classical HJ dynamics:
\begin{eqnarray}\left\{\begin{array}{ll}
-\partial_tS(q,t)=\sum_k \frac{1}{2m_k}(\partial_kS(q,t))^2+V(q,t)\\
m_k\frac{d}{dt}q_k(t)=\partial_kS(q(t),t)
   \end{array}\right.\label{27}
   \end{eqnarray}
with Eq.~\ref{26} we see very strong formal similarities. First of all, in both cases we have first-order systems of equations for the particle and guiding field dynamics $\Psi(q,t)$ and $S(q,t)$. Moreover, $\Psi(q,t)$ and $S(q,t)$ both act in 3$N$-dimensional configuration space (unlike the classical Hamiltonian, which acts in $6N$-dimensional phase space), so there is no feedback action of $q$ on $\Psi(q,t)$ and $S(q,t)$. Finally, the classical $S(q,t)$, or more precisely $e^{iS(q,t)}$, defines the semiclassical WKB limit of $\Psi(q,t)$. So if there's an analogy to be made to develop the nomological approach, it seems to us to be between the systems of relations \ref{26} and \ref{27}.\\
\indent But there's a big difference between \ref{26} and \ref{27}. Indeed, the meaning given by DGZ and Allori to the notion of nomological entity has always remained rather vague at the metaphysical or ontological level (even if their formulation is technically  clean).  But the classic \ref{27} system enables us to give a clear meaning to the expression “this is a nomological property and that is an ontological property”. Indeed, in the system \ref{27}, the $S(q,t)$ field is by no means necessary to describe the classical dynamics of the particle: it can be completely eliminated from the theoretical description (as clearly emphasized by Goldstein).
It can be replaced by the hydrodynamic equations: 
\begin{eqnarray}\left\{\begin{array}{ll}
\frac{d}{dt}p_k(q(t),t):=(\partial_t+\sum_j \frac{p_j}{m_j}\partial_j)p_k(q,t)=-\partial_kV(q,t)\\
\omega_{jk}(q,t)=\partial_jp_k-\partial_kp_j=0\\
p_k(q,t)=m_kv_k(q,t):=m_k\frac{d}{dt}q_k(t)
\end{array}\right.\label{28}
   \end{eqnarray}
The zero vorticity constraint $\omega_{jk}(q,t)=0$ implies of course that $v_k(q,t)=\frac{p_k}{m_k}=\frac{\partial_k S(q,t)}{m_k}$ and to solve the Euler-Newton equation we must choose as initial condition  a velocity field of zero vorticity, i. e., $v_k(q,t=0)=\frac{\partial_k S(q,t=0)}{m_k}$.   This shows that $S(q,t)$ is only a tool in this description, and of course in classical physics we can extend the field of possibilities by relaxing the zero vorticity constraint. If this constraint is eliminated, then Newton's equation becomes self-sufficient: it's no longer even necessary to speak of an Eulerian velocity field $v_k(q,t)$, just the Lagrangian or Newtonian description is sufficient. We have the fondamental second order law:
\begin{eqnarray}
m_k\frac{d^2}{dt^2}q_k(t)=-\partial_kV(q(t),t)\label{29}
   \end{eqnarray}  
However, the situation is completely different for the Bohmian system of equations \ref{26}. In de Broglie Bohm theory, the wave function $\Psi(q,t)$ cannot be eliminated.   More precisely, as we saw in the previous section, using the polar representation, the pair of equations \ref{26} actually becomes the triplet:
\begin{eqnarray}\left\{\begin{array}{ll}
-\partial_tS(q,t)=\sum_k \frac{1}{2m_k}(\partial_kS(q,t))^2+V(q,t)+V_\Psi(q,t)\\
\sum_k\partial_k(R^2(q,t)\frac{\partial_k S(q,t)}{m_k})+\partial_tR^2(q,t)=0\\
m_k\frac{d}{dt}q_k(t)=\partial_kS(q(t),t)
   \end{array}\right.\label{30}
\end{eqnarray}
In this set of equations \ref{30} the phase $S(q,t)$ and the modulus $R(q,t)$ are fundamentally entangled. In particular, the quantum potential $V_\Psi(q,t)$ depends on $R(q,t)$, which appears in the conservation equation of the probability fluid.     When we switch to the Newtonian quantum equation \ref{21} we see that the quantum potential is still there, of course, and so is the wave function $\Psi(q,t)$.  In other words, if we start with the hydrodynamic equation system:
\begin{eqnarray}\left\{\begin{array}{ll}
\frac{d}{dt}p_k(q(t),t):=(\partial_t+\sum_j \frac{p_j}{m_j}\partial_j)p_k(q,t)=-\partial_k(V(q,t)+V_\Psi(q,t))\\
\sum_k\partial_k(R^2(q,t)\frac{p_k}{m_k})+\partial_tR^2(q,t)=0\\
\omega_{jk}(q,t)=\partial_jp_k-\partial_kp_j=0\\
p_k(q,t)=m_kv_k(q,t):=m_k\frac{d}{dt}q_k(t)
\end{array}\right.\label{31}
   \end{eqnarray}
we can't reduce the dynamics to the Bohm-Newton equation \ref{Newton} alone, because the quantum potential requires the wave function by construction, and the probability density always remains coupled to $\Psi(q,t)$.\\
\indent In our view, this is a very serious objection to the nomological interpretation, which cannot be sustained.    
If we nevertheless wish to start from the system of equations \ref{26}, it is possible to construct a somewhat different interpretation by assuming that in de Broglie Bohm theory it is necessary to introduce as primitive variables (i.e., primitive ontology) both the positions $q(t)$ of the point particles in real 3D space and the wave function $\Psi(q,t)$ defined in configuration space. In other words, the variables $q$ and the wave function $\Psi$ (or its polar representation in terms of functions $R$ and S) are equally real and ontological.  This clearly imply a different primitive ontology not restricted to local beables $q$ but also including the wave function $\Psi$ defined in the configuration space and acting nonlocally on the particles. Moreover, such an approach treats the wave function in de Broglie Bohm theory as a physical entity with no real perfect classical analogues.  Such a so-called `sui generis' position is defended, for example, by the philosophers Chen \cite{Chen} and  Maudlin \cite{Maudlin} (who relies on the minimalist formulation of Bell \ref{26}), but also independently by the physicist Valentini \cite{Valentini}, who has been developing a competing interpretation to that of DGZ since 1992 and who defines the phase or action $S(q,t)$ as a real fundamental field.\\
\indent As an historical note, it is perhaps important to note that debates about the nomological or more ontological nature of the $\Psi$ wave function hark back to debates and discussions that took place in the 19th century about the physical meaning of gravitational and electrical potentials. Initially simple calculation tools and intermediaries in the work of Lagrange, Laplace and Poisson, they became independent objects in their own right in the work of Faraday and Maxwell. The big leap was actually made when physicists realized that potentials obey time-dependent differential equations. They thus became true independent physical variables (fields) on the same level as $q$ positional variables. The need for initial conditions for both the field and the particle showed that the field could not generally be eliminated, and that it was not simply a nomological tool. The analogy with the quantum debate is clear, and modern philosophers would do well to refer to the history of ideas in their metaphysical debates.\\
\indent Note that these ontological approaches come close to the `causal' interpretation proposed by Bohm and Hiley \cite{BohmHiley}, in which the wave function has no perfect classical analogue, making it impossible to really interpret the quantum potential $V_\Psi$ as a mechanical field in the classical sense of the word. According to them, it is a nonlocal information field (which they call `active information' to differentiate it from Shannon information) with no classical counterpart. This type of field cannot be compared to a signal, as it is not possible in the de Broglie Bohm theory to identify a source for the quantum potential. Note that some authors define the notion of `multi-fields' connecting several points defined in real space in order to talk about the wave function \cite{Hubert}. However, all these alternatives confirm the fundamentally new character of the wave function as a physical variable, i.e. to use John Bell's lexicon, as a `nonlocal beable'.\\
\indent However, perhaps to soften or weaken our point a little, we don't think that an extreme position that would consider saying that the $\Psi$ wave function is just something completely new is the right attitude either. Once again, this would mean overlooking the importance of de Broglie and Madelung's hydrodynamic formalism, which suggests a close kinship between Bohmian theory and classical physics. Moreover, we emphasize that despite some differences with classical forces the quantum potential is still a good guide for developping mechanical analogies (i.e., contrarily to stronger claims by Valentini, DGZ and Maudlin). On the other hand, we cannot reject in advance the idea that a better theory will one day replace quantum mechanics by restoring the priority of local variables in space-time. This was certainly Einstein's wish, but it was also that of de Broglie, who sought to develop his theory of the double solution exclusively in space-time and not the more abstract configuration space. In this context of the double solution \cite{debroglie1927,debroglie1956}, a return to the more traditional mechanical explanation is immediately essential, provided we can explain the presence of nonlocality. This could make sense in recent variants of the double solution that use retrocausality (see \cite{Wheeler,Wheelerb} for a classical electrodynamical model) as an alternative relativistic mode of explanation to Bohmian nonlocality \cite{Drezet1,Drezet2,Jamet}.\\
\indent To end this section, we think there's another argument in favor of the ontological (i.e. not just nomological) nature of the $\Psi(q,t)$ field. Indeed, we haven't mentioned the condition of continuity, regularity and single valuedness of $\Psi(q,t)$, which is however clearly implicitly presupposed in Eqs. \ref{26}, \ref{30} or \ref{31} and which implies the validity of the topological quantization relations \ref{kelvin1} and \ref{Kelvin}. As we indicated at the end of section \ref{section2}, these conditions are guided by physics and experience rather than mathematical necessity.  It should be noted that in the framework of classical wave optics based on Maxwell's equations, such quantization condition are observed and appear natural because the classical electromagnetic field is assumed to have a continuous ontological nature (in the classical framework it was a vibration of the Ether). Discontinuities in the field are generally admitted only in rapid or violent transient phenomena or during interaction with material interfaces. Moreover, rigorously discontinuities or dislocations are generally only approximation in optics and they always must fulfill the conditions \ref{kelvin1}, 
\ref{Kelvin}. In other words: Wave-fields like continuity.\\
\indent Returning to de Broglie-Bohm quantum mechanics, we think that what we've just said argues a little more strongly for the ontological nature of the $\Psi(q,t)$ field as a beable (even if nonlocal and despite important issues related to the physical meaning of empty waves \cite{debroglie1956,Selleri,Croca,Hardy,Vaidman,Drezet2006,Drezet2}). Note once again that the continuity postulate on $\Psi(q,t)$ is not true in general in classical physics. Importantly, in classical HJ theory, the $S(q,t)$ action can be multivalued, as in the case of classical particle scattering by Coulomb potentials generating $S$ discontinuities due to the presence of Caustics. In this example, `two Riemann sheets' are connected by a caustic to separate the incoming and outgoing solutions of the HJ equation and the associated part of the hyperbolic orbits. In real space, all incoming and outgoing trajectories can intersect (this type of phenomenon does not occur in de Broglie Bohm theory--see below). Similarly, in classical Eulerian hydrodynamics, the circulation of the velocity field around a vortex in a potential fluid is not quantized by an integer $N$ (moreover in hydrodynamics the velocity field is still regular and continuous). An often-commented-on feature of the de Broglie Bohm theory \cite{Holland} is that, unlike in the classical case (see \cite{Rowe}, for example), the trajectories of predicted particles cannot intersect in configuration space. The multi-valuedness of $S(q,t)$ in quantum physics (around phase singularities) does not affect the continuity of congruent trajectories in configuration space. This property leads to an apparent mismatch between classical and quantum theory in the regime of so-called 'surrealist' trajectories, where a Bohmian particle that from a classical point of view should pass from point A to point A' in fact ends its trajectory at point B', which again classically would be the logical termination of a trajectory originating from point B. This anti-crossing can be eliminated if we couple the Bohmian particle to an environment which, due to decoherence in a larger configuration space, allows the particles' trajectories to cross. Some of these counter-intuitive phenomena had already been discussed and analyzed by de Broglie and Brillouin at the Solvay Congress in 1927.  Note that, from a Newtonian point of view, it is the singularity induced by the quantum potential $V_\Psi(q,t)$ at the points where the wave function cancels out (which typically occurs in interference phenomena) that is responsible for the repulsive force between the Bohmian trajectories of the congruence. Once again, the absence of this potential allows the trajectories to cross each other in the classical regime.  All  these examples show important differences between classical $S$ action, or velocity potential, and its quantum version, i.e. Bohmian $S$ action.  In this context, we can observe that the interpretation of the wave function as a physical field is naturally advanced to explain the phenomena of superconductivity and superfluidity at very low temperatures, where a set of bosons are carried by the same $\Psi(q)$ wave at the lowest energy level. Quantum vortices are observed for example in superconductors (quantized  flux), and  Bose-Einstein condensates or polaritonic fluids, and the multivaluedness of the phase $S(q,t)$ is clearly interpreted using the Madelung formalism (see eg. Feynman \cite{Feynman}).  We believe that all this makes more sense if the field $\Psi(q,t)$ has a clear ontological character in line with what is admitted for classical waves (in optics or acoustics, for example) even if we must also consider  quantum nonlocality associated with $\Psi(q,t)$ as a completely new feature absent of classical ontology.

%%%%%%%%%%%%%%%%
\section{Can we complete or extend  Bohmian mechanics? The case of Born's rule} \label{section4}       
The previous section emphasized the new and original physical character of the wave function $\Psi(q,t)$ or quantum potential $V_\Psi(q,t)$ as a physical (ontological) variable or cause which, although defined in configuration space (or as a multifield), is indispensable to the description of the quantum system. However, not all advocates of de Broglie's theory agree on the meaning of $V_\Psi$. Some authors, such as DGZ \cite{DGZ1} and Valentini \cite{Valentini}, refrain from referring to this quantity at a fundamental level, as they believe it suggests too classical a mechanical explanation.   Others, such as Holland \cite{Holland}, Bohm and Hiley \cite{BohmHiley}, consider quantum potential to be a good explanatory tool that also provides a link with the classical world. Note that Bohm, Hiley and Holland do not suggest a return to classical ontology, as is clear from their use of the active information analogy, which emphasizes the non-mechanical character of $V_\Psi$. \\
\indent In the remainder of this article, we'd like to follow Holland's suggestions regarding the explanatory power of quantum potential. Indeed, the very motivation behind Bohm's theory is to be able to re-establish causality (hence the name causal interpretation attributed by Bohm). So we'll start from the premise that it's a pity to deprive ourselves of the mechanical and classical analogies that quantum potential and hydrodynamic formalism provide, even if we then have to deviate from them in the end.   The situation is similar to that of de Broglie, who, deeply rooted in history, sought to establish a link between Hamilton Jacobi's classical mechanics, geometrical optics and quantum physics \cite{debroglie1930}. \\
\indent Another central motivation for this analogy-based approach is the close relationship between de Broglie Bohm's pilot-wave theory and de Broglie's research into double-solution theory \cite{debroglie1927,debroglie1956}, which he began in 1923-25. In this approach, particles are assumed to be 'hump fields' or solitons, localized around the Bohmian trajectory and moving with it (the soliton is assumed to be guided by the phase of the $\Psi$ wave function). The double solution suggests a hydrodynamic analogy in which particles are extended objects immersed in a quantum fluid that Bohm and Vigier named the subquantum medium \cite{Vigier,Vigierthesis,BohmHiley}. In this context, Madelung de Broglie's hydrodynamic analogies take on their full meaning and merit further study.\\
\indent In the same vein, over the last two decades a great deal of theoretical and experimental work has focused on the hydrodynamic analogies suggested by Couder-Fort and Bush's experiments with bouncing drops or walkers \cite{Bush,Couder}. In these purely classical systems, drops bounce off an oil bath oscillating at a constant frequency, generating waves that guide the drop as it moves over the bath. The interaction of these waves with obstacles and the immediate environment retroacts on the drop, suggesting de Broglie-Bohm-type guiding behavior. This defines a framework that is also of great interest to Madelung-de Broglie hydrodynamic analogies.\\
\indent The central point, in our view, is that de Broglie-Bohm theory, by proposing an explanatory and causal model for quantum mechanics, also allows us to glimpse potential strategies for going beyond or complementing it. This aspect has been often underlined by the creators of this theory since de Broglie, Bohm and Vigier \cite{debroglie1956,Vigier,Vigierthesis}, who envisaged the pilot wave theory as a guide to obtaining a theory that would go beyond quantum mechanics and perhaps enable a better understanding of the theory of elementary particles (the quantum theory of relativistic fields) and also make a link with Einstein's theory of general relativity for a possible unification of quantum physics and gravitation (which is still in the sketch stage). \\
\indent To be more explicit in what follows, we would like to consider different possibilities for extending or completing the Bohmian formalism, based on the hydrodynamic analogies suggested by the Madelung and de Broglie formalism (cf. Eq. \ref{31}) and relaxing or modifying some of the fundamental assumptions. More specifically, we will consider two possibilities, concerning either i) the notion of probability and Born's law $\rho(q,t)=R^2(q,t)$, or ii) dynamics and the irrotational postulate $\omega_{ij}(q,t)=0$. In this section we'll briefly describe and review possibility i), which has already been discussed at length in the literature and is the source of much controversy. Possibility ii) will be discussed in the following sections.\\
\indent The idea of relaxing Born's rule $\rho(q,t)=R^2(q,t)$ dates   back to the early work of de Broglie \cite{debroglie1926b,debroglie1927}, who from 1926-27 understood that to recover quantum mechanics and its statistical prediction, it was necessary to impose this formula in the postulates. In fact, de Broglie introduced the statistical formulation in a short 1926 paper focusing on photons \cite{debroglie1926b}. In 1927, de Broglie generalized his postulate to all particles, after becoming aware of Born's theory\cite{debroglie1927,debroglie1927b}.    For de Broglie, unlike Born, probabilities are not fundamental and must be derived from additional postulates about the initial conditions of particles.  These initial conditions, and therefore the statistical postulate, are much more contingent in nature than they were for Born and the Copenhagen school. Inspired by the masters of 19th-century statistical physics, De Broglie imagines that particles are initially distributed in such a way as to satisfy Born's rule, and demonstrates that if this is true at one instant, it will be true at all times. More precisely, let's assume that the conservation law $\sum_k\partial_k(R^2(q,t)\frac{p_k}{m_k})+\partial_tR^2(q,t)=0$ in which $|\Psi|^2:=R^2$ is associated with the guiding field is true, but is not necessarily a probability density.   According to what we said in section \ref{section3}, the $\Psi(q,t)$ field is indeed a fundamental physical property and cannot be reduced to a probability field.    We therefore need to develop a second conservation law to deal with the distribution of particles in configuration space, assuming that the universe contains a large number of identically prepared copies in the same $\Psi$ and $V$ fields, but with different $q(0)$ initial conditions. The local conservation law for this fluid is written in all generality (as in classical physics):
\begin{eqnarray}
\sum_k\partial_k(\rho(q,t)\frac{p_k}{m_k})+\partial_t\rho(q,t)=0
\end{eqnarray}
with in general $\rho(q,t)\neq |\Psi|^2(q,t)$.   By comparing  with the local conservation law for $R^2$ we deduce the relation 
\begin{eqnarray}
[\partial_t+\sum_k\frac{p_k}{m_k}\partial_k](\frac{\rho(q,t)}{R^2(q,t)}):=\frac{d}{dt}(\frac{\rho(q,t)}{R^2(q,t)})=0
\end{eqnarray}
which shows that the function $f(q,t):=\frac{\rho(q,t)}{R^2(q,t)}$ is an integral of the motion along the trajectory of the system $q(t)$ \cite{Vigier,Vigierthesis,debroglie1956}. Basically, this formula is very similar to Liouville's theorem in classical statistical physics (although the latter is written in phase space and not configuration space). Indeed, if we write $\delta P(q,t)$ the probability of presence in the configuration space element $\delta^{3N}q$ we have 
\begin{eqnarray}
\delta P(q,t)=\rho(q,t)\delta^{3N}q=f(q,t)\delta \Gamma
\end{eqnarray}
 where the quantity $\delta \Gamma:=|\Psi|^2(q,t)\delta^{3N}q$ plays the role of a measure conserved in configuration space (we have $\frac{d}{dt}\delta \Gamma(q(t),t)=0$). Since, by conservation of the probability fluid, we must also have $\frac{d}{dt}\delta P(q(t),t)=0$, we find again the relation $\frac{d}{dt}f(q(t),t)=0$, which is exactly the same derivation as for Louville's theorem and allows us to interpret $f(q,t)$ as a probability density with respect to the $\Gamma$ measure in configuration space.\\
 \indent  In this formalism, de Broglie's statistical postulate, which we'll call the quantum equilibrium postulate, is to impose $f=1$ everywhere at time $t=0$, knowing that this will remain true for all times $t$ along the Bohmian trajectories $q(t)$. This postulate resembles the micro-canonical  equilibrium postulate in statistical physics, but once again it is obtained in $3N$-dimensional configuration space, not $6N$-dimensional phase space, and the $d\Gamma$ measure plays an essential role, replacing the $d\Gamma_{phase}:=d^{3N}qd^{3N}p$ measure in phase space in ordinary statistical physics. The problem, of course, is that even if this microcanonical postulate seems natural, since it introduces a form of equiprobability with respect to the $\Gamma$ measure in configuration space, it is nonetheless contingent and in no way necessary to the dynamic theory of de Broglie Bohm. Other choices are possible: for example we could consider $f$ constant and different of zero only in a trajectory tube and $f=0$ outside.  Ultimately,  we could also choose $f(q,t)=c\delta^{3N}(q-q_0(t))/|\Psi(q,t)|^2$ where $q_0(t)$ is a particular  Bohmian trajectory and $c$ a normalization constant (it can be checked that this distribution satisfies the constraint $\frac{d}{dt}f(q(t),t)=0$). Since then, this point concerning microcanonality has been criticized, notably by Pauli and Keller \cite{Pauli1953,Keller1953}, as arbitrary.\\
 \indent In response to such criticisms, Bohm Vigier and de Broglie \cite{Vigier,Vigierthesis,debroglie1956} defended the idea that the $f=1$ microcanonical distribution corresponding to quantum equilibrium would be an attractor state due to the chaotic complexity of a system's Bohmian interactions with its environment.  Inspired by Boltzmann's strategies, they tried to demonstrate that the quantum entropy $-\int d\Gamma f\log{f}$ naturally relaxes to a maximum for the state $f=1$. To this end, they also introduced the hypothesis of molecular chaos in a hypothetical sub-quantum medium interacting with the system under consideration.\\
 \indent Much later, Valentini \cite{Valentini,Valentini2020} took up Bohm-Vigier's program, but eliminated the too hypothetical sub-quantum medium and replaced the Boltzmanian approach with that of Gibbs, using coarse-grained averages to average the information and thus create an entropy-increasing mixture. The ultimate aim was still to justify the approach to quantum equilibrium, this time invoking a loss of information due to our approximation in the treatment of the complexity of Bohmian particle motion. Valentini and his colleagues demonstrated with the help of numerous numerical examples \cite{Valentini2005,Valentini2012} that a density initially based in a non-equilibrium state $f=1$ tends rapidly in general towards the homogeneous state $f=1$. This argument, which resembles Gibbs' in classical statistical physics here, works essentially because of the highly nonlinear nature of de Broglie Bohm mechanics (in particular, because of the complexity of the quantum potential $V_\Psi(q,t)$).  An alternative (advocated by one of the present authors \cite{DrezetEntropy,Drezet2017}) is to connect the information loss associated with Valentini coarse-grained averages to the quantum entanglement of a subsystem with a thermal bath (thermostat) already assumed to be in a state of quantum equilibrium. The idea is that, due to quantum coupling, a subsystem initially in a state very different from $f=1$ will be attracted (relaxation) to the microcanonical $f=1$ state. This mode of relaxation is not independent of Valentini's approach.\\
 \indent It's important to note that no deviation from quantum equilibrium, i.e. from Born's rule, has ever been observed experimentally. Clearly, if this were to happen, the consequences would be far-reaching. For example, as Valentini was quick to point out \cite{Valentini1991}, due to the nonlocality of Bohmian mechanics, any deviation from the $f=1$ law could in principle be used to send a signal faster than light. This is, of course, in violation of the non-signalling theorem used in the discussion of Bell's theorem, which prohibits such effective supraluminal communication. However, this theorem is based, among other things, on the validity of Born's rule, and its weakening would lead to new physics. An original possibility, for example, would be to exploit measurements of arrival times predicted by Bohmian theory \cite{Das,DrezetDas} in a quantum non-equilibrium regime.  Valentini's current hope \cite{Valentini2007,Valentini2015,Valentini2020} is that such violations may one day be detectable via fluctuations in primordial cosmic radiation that cannot be explained by Born's rule (this presupposes a state of primordial quantum non-equilibrium). This research is of course very interesting and disserves further studies. \\
 \indent We can't end this section without mentioning the competing approach to that of Valentini Bohm and Vigier, advocated by DGZ and  which posits Born's law as somehow inevitable and natural \cite{DGZ1992}. This approach is based on the Boltzmanian notion of typicality and assumes that for the universe system taken as a whole, the $|\Psi(Q,t)|^2$ distribution (where $Q$ takes into account all particles) is not associated with   a statistical distribution (because there is only one Universe). However, according to DGZ, $|\Psi(Q,t)|^2d^{3N}:=d\Gamma$ is still the most natural probability measure for describing configuration space, as it is equivariant. Indeed, if we are looking for a distribution $\rho$ that is an explicit function of $|\Psi|^2$, i.e. $\rho=F(|\Psi|^2)$, then the only admissible solution is $F(x)=x$, i.e. $\rho=|\Psi|^2$ and this property is preserved with time: if $\rho_{t_0}=|\Psi_{t_0}|^2$ holds true at a given time it will be true at any other times $\rho_{t}=|\Psi_{t}|^2$. The $d\Gamma$ measure is therefore apriori the obvious choice to weight the configuration space of the  Universe. To give a physical meaning to the weight or measure $d\Gamma$ we need to consider a very large set of quantum-identical subsystems. In other words, we assume that our universe has $N$ factorizable subsystems described by the same wave function $\psi (q_i,t)$ ($3M$ is the dimension of each $i$ subsystem and $q_i\in \mathbb{R}^{3M}$ ). For these $N$ subsystems, we then have a global effective wave function $\psi(q_1,t)\psi(q_2,t)...\psi(q_N,t)$. Applying the law of large numbers (Bernoulli) to this long sequence of systems weigthed by the $\Gamma$ measure shows that with high probability the statistical distribution $\frac{1}{N}\sum_i \delta^{3M}(q-q_i(t))$ (with $q\in \mathbb{R}^{3M}$) tends towards $|\psi(q, t)|^2$. The word very probable, or typical, stands for ``almost every systems", i.e. it implies in the overwhelming majority of cases the empirical validity of Born's rule at the statistical level for sufficiently large samples ($N>>1$). Fluctuations to the rule become negligible, i.e. atypical.\\
 \indent The consensual part of DGZ's deduction concerns the application of the law of large numbers, which is also used by Valentini in his deductions. In our view, there are two points of criticism in DGZ's analysis. The first is semantic, and the least important, although it is a source of confusion. DGZ and their colleagues \cite{DGZ1992,Teufel,Durr} use the term typicality measure instead of the more common term probability used in the literature. This is linked to a historical confusion: contrary to what they claim, a probability is not a relative statistical frequency; it's the tool invented by the masters of probability theory, from Bernoulli to Laplace, Borel and Kolmogorov, which enables us to reproduce these statistical frequencies. There's no need to introduce a new notion, typicality, to replace the more accepted notion of probability. Moreover, the point that has generated the most debate concerns the need to use the equivariant measure $d\Gamma$, which in fact amounts to postulating from the outset the microcanonality or equiprobablity $f=1$ for the Universe. In fact, despite its logical simplicity, there's nothing to impose the equivariance rule as the most natural. We have seen that the equally elegant distribution $f(q,t)=c\delta^{3N}(q-q_0(t))/|\Psi(q,t)|^2$ could be chosen by Nature but is clearly not!  The degree of contingency of a probability law cannot be eliminated by decree, and only physics (i.e. experience) decides what is typical and what is not. In other words, the quantum equilibrium rule, being contingent, is still not, despite DGZ's elegant reasoning, a law of nature, and in fact its justification always seems to be deferred to a particular choice of initial conditions of the Universe (in connection with cosmology). The merit of Bohm Vigier and Valentini's program, in our view, is that it offers the promise of a better understanding of the robustness of this Born rule (via coupling with a thermostat, for example). 
%%%%%%%%%%%
\section{Dynamical completion of the Bohmian dynamics: Bohm's proposal and its critics }\label{section5}
As we have seen, many Bohmians (DGZ \cite{GoldsteinTumulkaZanghi}, Valentini \cite{Valentini}) have strongly criticized Bohm's formulation in terms of quantum potentials $V_\Psi$ and his use of Newton's law. Quantum potential suggests highly complex and, in the words of DGZ and Valentini, inelegant dynamics. Moreover, apart from the criticisms already made, it seems that quantum potential suggests a coupling force between trajectories (this is the analogy with Madelung's tension field, which presupposes an interaction force between several fluid elements, i.e. several trajectories). In this case, however, there are not several systems at once: a single Bohmian trajectory is realized, and it seems strange to see an influence of the collective on the individual in this Bohmian theory. \\
\indent This curious impression, already noted by Kennard and Rosen \cite{Kennard,Rosen}, is reinforced if we express $V_\Psi$ in the hydrodynamical form $V_\Psi(q,t)=\sum_k \frac{-1}{2m_k}\frac{\partial^2_k \sqrt{\rho}(q,t)}{\sqrt{\rho}(q,t)}=\sum_k \frac{-1}{4m_k}[\frac{\partial^2_k \rho(q,t)}{\rho(q,t)}-\frac{(\partial_k\log{\rho(q,t)})^2}{2}]$, which brings out the (quantum equilibrium) probability density $\rho=R^2$.  In fact, most of these criticisms are unjustified. For example, similarity with a real fluid does not necessarily imply the existence of several simultaneously existing trajectories (this is not so in the classical HJ formalism), nor should the density $R^2(q,t)$ be fundamentally interpreted as a probability density: De Broglie, Kennard Rosen and Bohm are very clear on this point. The wave function has an ontoplogical status not an epistemic one and $R$ is before all a dynamical variable not the square root of $\rho$.  In fact, the rejection of quantum potential seems more a matter of taste than of substance.\\
\indent If we accept this point, then we can follow Bohm in some of his suggestions: that's what we'll do from now on.  Rigorously speaking, Bohm  \cite{BohmI} considers the system of equations \ref{31} to be the starting point, and so the zero vorticity condition $\omega_{ij}(q,t)=0$ plays an essential role. It defines a constraint on quantum Newton dynamics, and this constraint is specifically Eulerian in nature, i.e. it is defined on the velocity field in configuration space.   In other words, the relation $p_k=\partial_k S(q,t)$ defines a subset or class of Bohmian trajectories obeying the Newton-Bohm law and subject to the constraints given in  \ref{30} or \ref{31}. Bohm in 1952 \cite{BohmI} clearly suggested that condition of null vorticity could be relaxed, i.e. that we could perhaps abandon the guiding condition $p_k=\partial_k S(q,t)$. However, it appears that there are at least two ways of understanding this abandonment of constraint.   Bohm's version, which we'll briefly outline, and Sch\"onberg's more interesting one, which we discuss in the next section.\\
\indent Bohm's suggestion is to start from the new system 
\begin{eqnarray}\left\{\begin{array}{ll}
i\partial_t\Psi(x,t)=\sum_k \frac{-1}{2m_k}\partial^2_k \Psi(x,t)+V(x,t)\Psi(q,t)\\
m_k\frac{d^2}{dt^2}q_k(t):=-\partial_k(V(q(t),t)+V_\Psi(q(t),t)
   \end{array}\right.\label{35}
   \end{eqnarray}
instead of \ref{26} and see how these relations can be modified, either by adding nonlinear terms to the Bohm-Newton equation or to the Schrodinger equation, in order to induce a kind of forced relaxation towards the guiding condition $m_k\dot{q}_k(t)=p_k(q(t),t)=\partial_kS(q(t),t)$.   
Let's note that \ref{35} amounts to adding Newton's relation to the hydrodynamic equations \ref{30} or \ref{31} deprived of the guiding formula $m_k\dot{q}_k(t)=p_k(q(t),t)=\partial_kS(q(t),t)$. In fact, to be more precise in this theory, we can always define a Lagrangian fluid velocity in accordance with Lagrange's definition of a Madelung fluid, i.e. $p_k(x(t),t)=\partial_kS(x(t),t)=m_k\dot{x}_k(t)$ with $x_k(t)$ a fluid trajectory. However, this fluid trajectory (path line of a `fluid molecule') is no longer generally identifiable in Bohm's \ref{35} approach with the actual particle trajectory $q_k(t)\neq x_k(t)$.\\
\indent The problem is that this new dynamic is much richer than the original de Broglie-Bohm theory, because the Newton-Bohm law $m_k\frac{d^2}{dt^2}q_k(t):=-\partial_k(V(q(t),t)+V_\Psi(q(t),t)$ contains many solutions that contradict quantum mechanics and make the theory unstable and not very credible. For example, as discussed in \cite{Colin}, according to this theory, an electron in the ground state of the hydrogen atom is not subject to any force because the repulsive quantum potential rigorously balances the attractive Coulomb potential. Therefore, according to Newton Bohm's law $m\frac{d^2q(t)}{dt^2}=0$ the electron's motion can in such an atom ground state  follow an inertial motion at constant speed and could therefore escape to infinity. Of course this is not so in the de Broglie Bohm theory where the electron is at rest. Furthermore according to this theory and for the same atomic example, we can construct  an unrealistic statistical ensemble with particles all having a velocity in the $z$ direction: the Eulerian velocity field in this statistical ensemble is given by an arbitrary function of $x,y$, i.e. $\mathbf{v}(x,y,z)=U(x,y)\mathbf{\hat{z}}$ (and should not be confused with the Eulerian velocity field deduced from the wave function which predicts $\mathbf{v}_{dBB}=\boldsymbol{\nabla}S/m=0$). We can introduce a  probability density $\rho(x, y,z)=F(x,y)$ given by an arbitrary function of $x$ and $y$ (we have local probability conservation $\boldsymbol{\nabla}\cdot(\rho \mathbf{v})=\partial_z(FU)=0$). This distribution is not confined in the atom potential which demonstrates the apriori non-physical nature of this model. Further numerical calculations confirm the unstability of Bohm's proposal \cite{Colin,Struyve}. Hence, without the addition of nonlinear terms (of unknown precise form) in \ref{35} forcing the particle system to converge towards de Broglie Bohm trajectories obeying the guidance formula $m_k\dot{q}_k(t)=p_k(q(t),t)=\partial_kS(q(t),t)$, it is impossible to accept such an approach.\\
\indent Another way of looking at the problem is to consider a Hamiltonian formulation instead of the Newtonian one. In this approach, we replace Eq.\ref{35} by the equivalent system 
\begin{eqnarray}\left\{\begin{array}{ll}
i\partial_t\Psi(x,t)=\sum_k \frac{-1}{2m_k}\partial^2_k \Psi(x,t)+V(x,t)\Psi(x,t)\\
\dot{q}(t)=\frac{\partial H_\Psi(q(t),p(t))}{\partial p}\\
\dot{p}(t)=-\frac{\partial H_\Psi(q(t),p(t))}{\partial q}
   \end{array}\right.\label{36}
   \end{eqnarray}
which contains the Hamiltonian $H_\Psi(q,p,t)=\sum_k \frac{1}{2m_k}p_k^2+V(q,t)+V_\Psi(q,t)$ with here $p_k$ and $q_k$ the canonical variables associated with the particles, not to be confused with the variables obtained from the Madelung formalism and associated with the wavefunction [here the distinction between the positions variables $x$ associated with $\Psi$ and the variables $q$ associated with the particles is particularly useful]. The problem is similar to what happens in the Newtonian formulation: the congruence of trajectories in the $6N$-dimensional phase space obtained from the Hamiltonian $H_\Psi(q,p,t)$ defines a much larger set than that of simple Bohmian trajectories subject to the $m_k\dot{q}_k(t)=p_k(q(t),t)=\partial_kS(q(t),t)$ constraint. If, for example, we are interested in the problem of quantum equilibrium, it is natural in this theory to start from the probability density $\eta(q,p,t)$ in phase space and it is well known that this distribution obeys Liouville's theorem 
\begin{eqnarray}
[\partial_t +\sum_k \dot{q}_k\frac{\partial}{\partial q_k}+\dot{p}_k\frac{\partial}{\partial p_k}]\eta(q,p,t):=\frac{d}{dt}\eta(q(t),p(t),t)=0
\end{eqnarray} 
However, this theorem admits several apriori possible solutions to define the notion of equilibrium, and the simplest would seem to be the microcanonical equilibrium $\eta(q,p,t)=\eta_0=const.$ which is not observed experimentally.
Other possibilities are of course possible, for example: 
\begin{eqnarray}
\eta(q,p,t)=c\delta^{3N}(q-q_0(t))\delta^{3N}(p-p_0(t))\label{37a}\\
\eta(q,p,t)=f(q,t)|\Psi(q,t)|^2\delta^{3N}(p-\partial_q S(q,t))\label{37b}\\
\eta(q,p,t)=|\Psi(q,t)|^2\delta^{3N}(p-\partial_q S(q,t))\label{37c}
   \end{eqnarray}
Eq.~\ref{37a} corresponds to a single trajectory  $q_0(t), p_0(t)$ and a state of strong quantum non-equilibrium  Eq.~\ref{37b} and Eq.~\ref{37c}  are close to Vigier and Valentini's formulation discussed before, where the wave function plays its role as an attractor. Only Eq.~\ref{37c} corresponds to quantum equilibrium with $f(q,t)=1$ (see \cite{Takabayasib,debroglie1956}). Bohm's generalized theory \ref{36} in phase space thus creates a much larger inventory of possibilities than the usual de Broglie Bohm version, and thus apriori adds problems to the existing formulation rather than solving them. Indeed, only the addition of extra terms (Bohm suggests expressions of the type $G[p-\partial_q S(q,t)]$) in the fundamental equations
\begin{eqnarray}\left\{\begin{array}{ll}
i\partial_t\Psi(x,t)=\sum_k \frac{-1}{2m_k}\partial^2_k \Psi(x,t)+V(x,t)\Psi(q,t)+G_1[p-\partial_x S(x,t)]\\
m_k\frac{d^2}{dt^2}q_k(t):=-\partial_k(V(q(t),t)+V_\Psi(q(t),t)+G_2[p-\partial_q S(q,t)]
   \end{array}\right.\label{35b}
   \end{eqnarray}
could potentially force motions to stick to Broglie Bohm trajectories guided by the wave function: all of which remains rather speculative. It should be noted that, despite these shortcomings, this in no way prejudices Bohm's general project, which was very much inspired by F\"urth's stochastic mechanics of the 1930s \cite{Furth}, and which attempted to establish a link between Heisenberg's uncertainty principle and the Brownian stochastic motion of a particle subjected to fluctuating forces. The idea inspired Bohm who, in collaboration with Vigier \cite{Vigier} and later Hiley \cite{BohmHiley}, introduced stochastic fluctuating terms into de Broglie Bohm dynamics to force convergence to quantum equilibrium over long time scales. Numerous other works flourished in the 1950s and 60s around stochastic quantum mechanics (the most famous being Nelson's theory \cite{Nelson} see also \cite{Comisar,Kershaw,Harvey}), most of them attributing an important role to the `osmotic velocity' \cite{Furth}, which for a particle is written as $\mathbf{v}_{osm}= D\boldsymbol{\nabla}(\log{\rho})$ where $D=\frac{\hbar}{2m}=\frac{1}{2m}$ is a quantum diffusion constant characterizing the `subquantum medium' \cite{Vigier}.

One criticism of this project may be that it departs sharply from de Broglie's initial desire to re-establish causality and determinism (even though de Broglie later supported the introduction of stochastic aspects into his double solution with his thermodynamics of the isolated particle). Moreover, the possible extension of the stochastic approach to the multi-particle problem poses new fundamental problems, since fluctuating quantum forces are also non-local, which sets them apart from the classical stochastic forces associated with local collisional processes.  Interestingly, DGZ and Tumulka \cite{QFT}, following an idea of Bell \cite{BellQFT} and Vink \cite{Vink}, have more recently reintroduced stochastic dynamics and nonlocality into the Bohmian version of quantum field theory (QFT) to account for the creation and disappearance of particles in these theories (Goldstein in particular studied Nelson's theory in detail in the 1980s \cite{Goldstein1987}).
%%%%%%%%%%%%%%%%%%
%%%%%%%%% 
\section{Extension of the de Broglie-Bohm framework involving vorticity and Clebsch potentials}\label{section6}
In order to motivate the present discussion, it is important to recall how HJ theory is generally introduced and justified in classical mechanics \cite{Holland,Landau,GoldsteinB}.    Starting from a Hamiltonian $H(q,p,t)$ we assume the existence of a canonical transformation, $q,p\rightarrow Q,P$ such that with the new variables the Hamiltonian $H'(Q,P,t)$ cancels. Such a canonical transformation is introduced, for example, by means of a generating function $F(q,Q,t)$ such that $p=-\frac{\partial F}{\partial q}$, $P=-\frac{\partial F}{\partial Q}$, and $H'=H+\partial_tF$. If $H'=0$ we have, according to Hamilton's equations $\dot{Q}=0=\dot{P}$, i.e. $Q$ and $P$ define constants of motion. Let $\alpha:= Q$, $S(q,\alpha,t):=F(q,Q,t)$ and $\beta=P$, we have 
\begin{eqnarray}
-\partial_t S(q,\alpha)=H(q,\nabla S(q,\alpha,t), t)\label{HJ1}\\ p=\nabla S(q,\alpha,t) \label{HJ2}\\
\beta=-\frac{\partial S(q,\alpha,t)}{\partial \alpha}\label{HJ3}
\end{eqnarray} Following Jacobi's theorem the action $S(q,\alpha)$ define a complete integral of the HJ equation \ref{HJ1} with the momenta $p_k=m_k \dot{q}_k$ given by the guidance formula \ref{HJ2}. This solution is characterized by  $3N$  non-additive constants $\alpha$ defining a family of trajectories in the configuration space. The other $3N$ constants of integration $\beta$ given by \ref{HJ3} characterize each specific orbit or trajectory of the family requiring   $6N$ integration constants for their complete definition.\\
\indent However, there are a number of points to bear in mind here. Firstly, Jacobi's method and result are much more general than our brief description based on a specific choice of generating function. Other choices of generating functions, such as $F_2(q,P,t)$, $F_3(p,Q,t)$ or $F_4(p,P,t)$, would be just as suitable for reaching the same general conclusion.  Another historical remark is that de Broglie was quick to appreciate that the existence of families of trajectories associated with the $S(q,t)$ function meant that a statistical element was introduced into the pilot-wave theory he was proposing. Traces of this can be found as early as 1925 \cite{debroglie1924,debroglie1925b,debroglie1926b,Drezet2025}, when he anticipated Born's rule even though the Schr\"odinger equation had not yet been discovered!   For this reason, we can truly say that pilot-wave theory, i.e. Bohmian mechanics, preceded the Copenhagen interpretation.\\ 
\indent A more critical remark is that there's no guarantee that we can always find a complete integral $S(q,\alpha,t)$ solution to the equation HJ \ref{HJ1}. The method often used is separation of variables, but it doesn't always work \cite{Holland}.  In fact, it seems preferable to look at the problem the other way round.  Postulating the HJ equation \ref{HJ1}, which is a nonlinear equation dependent on a field $S(q,t)$, we can always define a class or family of trajectories using the guiding formula \ref{HJ2}: this is the central postulate associated with \ref{27} in classical mechanics  and \ref{30} in Bohmian mechanics. All this becomes clearer if we start from the hydrodynamic formalism. In fact, the HJ method in classical mechanics is merely a means of describing the dynamics of a set of trajectories in configuration space, based on the analogy with a fluid without pressure $P$ (the only forces coming from the external field $V(q,t)$) as visible from Eq.~\ref{28}.  The HJ description is limited to potential motion, i.e. postulating irrotationality $\omega_{ij}(q,t)=0$. This is a restricted class of motion obeying Euler's equation, i.e. Newton's. The same is of course true in de Broglie Bohm mechanics (see Eq. \ref{31}) with the difference that in this quantum theory there is a Madelung-de Broglie internal tension field necessary to explain the quantum forces associated with $V_\Psi(q,t)$.\\ 
\indent This immediately suggests possible generalizations in both classical and Bohmian quantum physics. Indeed, if we remove this restriction to potential motions $\omega_{ij}(q,t)=0$ it becomes possible to obtain a much larger class of sets of trajectories in configuration space, containing those described by the HJ equation as a special case.   Such an approach is well known in hydrodynamics for inviscid fluids \cite{Scholle} and electromagnetics (e.g. magnetohydrodynamics) \cite{Stern,Merches}, where Clebsch potentials are introduced to describe velocity fields with vorticity.  It then becomes possible to imagine extensions of Madelung's formalism to fluids with non-zero vorticity fields. As demonstrated by Sch\"onberg and Takabayasi \cite{Schonberg,Takabayasi}, it even becomes possible to generalize Schr\"odinger's equation to take account of such non-potential particle motions.   Remarkably, a relativistic extension to the Klein-Gordon equation was independently proposed by Dirac \cite{Dirac} in the limit of zero quantum potentials, and its `Bohmian' generalization was proposed by Takabayasi and Sch\"onberg \cite{Takabayasi,Schonberg} (for the Klein-Gordon equation and Dirac's equation for $1/2$ spin particles).\\
\indent In the following, we will restrict ourselves to the nonrelativistic regime in a configuration space of dimension 3 ($N=1$) in order to simplify the analysis, although Clebsch potentials can in principle be discussed for $N>1$.  We start with  
\begin{eqnarray}\left\{\begin{array}{ll}
(\partial_t+\mathbf{v}(\mathbf{r},t)\cdot\boldsymbol{\nabla})\mathbf{v}(\mathbf{r},t)=-\boldsymbol{\nabla}\frac{V(\mathbf{r},t)+V_\Psi(\mathbf{r},t)}{m}\\
\partial_t R^2(\mathbf{r},t)+\boldsymbol{\nabla}\cdot(R^2(\mathbf{r},t)\mathbf{v}(\mathbf{r},t))=0
\end{array}\right.
\label{Cleb1}
\end{eqnarray} without Madelung's constraint  $\boldsymbol{\Omega}(\mathbf{r},t)=\boldsymbol{\nabla}\times\mathbf{v}(\mathbf{r},t)=0$. Following Euler we can always write in a region where $\boldsymbol{\Omega}(\mathbf{r},t)\neq 0$ 
\begin{eqnarray}
m\boldsymbol{\Omega}(\mathbf{r},t)=\boldsymbol{\nabla}{\alpha}(\mathbf{r},t)\times\boldsymbol{\nabla}{\beta}(\mathbf{r},t)\label{Cleb2}
\end{eqnarray} where $\alpha$ and $\beta$ are the so called Euler or Clebsch potentials. This representation can be justified by using Pfaff-Darboux's theorem \cite{Scholle,Clebsch,Hankel}. Comparing with the definition for $\boldsymbol{\Omega}(\mathbf{r},t)$ this leads to the Clebsch decomposition
   \begin{eqnarray}
m\mathbf{v}(\mathbf{r},t)=\boldsymbol{\nabla}S(\mathbf{r},t)+{\alpha}(\mathbf{r},t)\boldsymbol{\nabla}{\beta}(\mathbf{r},t)\label{Cleb3}
\end{eqnarray} which is however not unique since the following gauge  transformations are allowed \cite{Scholle}: 
\begin{eqnarray}
S\rightarrow S'=S+f(\alpha,\beta,t) \nonumber\\
\alpha\rightarrow \alpha'=g(\alpha,\beta,t)\nonumber\\
\beta\rightarrow \beta'=h(\alpha,\beta,t)
\end{eqnarray} if $\frac{\partial f}{\partial \beta}+g\frac{\partial h}{\partial \beta}=\alpha$, $\frac{\partial f}{\partial \alpha}+g\frac{\partial h}{\partial \alpha}=0$. Using the Lagrangian description of a fluid Clebsch and Hankel  (see \cite{Clebsch,Hankel} for a clear discussion of the historical proofs but also \cite{Stern} where the connection with Euler's work is done) showed that for consistency the potentials obeys the conditions 
\begin{eqnarray}
(\partial_t+\mathbf{v}(\mathbf{r},t)\cdot\boldsymbol{\nabla})\alpha(\mathbf{r},t):=\frac{d}{dt}\alpha(\mathbf{r}(t),t)=0\nonumber\\
(\partial_t+\mathbf{v}(\mathbf{r},t)\cdot\boldsymbol{\nabla})\beta(\mathbf{r},t):=\frac{d}{dt}\beta(\mathbf{r}(t),t)=0\label{Cleb4}
\end{eqnarray} 
It must be noted that Eq.~\ref{Cleb3} is looking very similar to the HJ guidance formula for a charged particle in an external magnetic potential $\mathbf{A}(\mathbf{r},t)$ reading 	$m\mathbf{v}(\mathbf{r},t)=\boldsymbol{\nabla}S(\mathbf{r},t)-e\mathbf{A}(\mathbf{r},t)$ where $e$ is the electric charge of the particle.\\
\indent Although we have not yet introduced electromagnetic fields into our discussion of the de Broglie Bohm theory, we recall that for a non-relativistic spinless particle obeying the Schrödinger equation in the presence of a magnetic potential $\mathbf{A}$ and an electric potential $V$ we have:  
\begin{eqnarray}
i\partial_t\Psi(\mathbf{r},t)=\frac{-1}{2m}(\boldsymbol{\nabla}-ie\mathbf{A}(\mathbf{r},t))^2\Psi(\mathbf{r},t)+eV(\mathbf{r},t)\Psi(\mathbf{r},t)\label{Cleb5}
\end{eqnarray}
 This corresponds to Madelung's hydrodynamic representation
 \begin{eqnarray}\left\{\begin{array}{ll}
-\partial_tS(\mathbf{r},t)=\frac{(\boldsymbol{\nabla}S(\mathbf{r},t)-e\mathbf{A}(\mathbf{r},t))^2}{2m}+V_\Psi(\mathbf{r},t)+eV(\mathbf{r},t)\\
\partial_t R^2(\mathbf{r},t)+\boldsymbol{\nabla}\cdot(R^2(\mathbf{r},t)\mathbf{v}(\mathbf{r},t))=0
\end{array}\right.
\label{Cleb6}
 \end{eqnarray}
which together with the guidance formula $m\mathbf{v}(\mathbf{r},t)=\boldsymbol{\nabla}S(\mathbf{r},t)-e\mathbf{A}(\mathbf{r},t)$ defines the associated Bohmian mechanics. The second order Newton's dynamics reads 
\begin{eqnarray}
m(\partial_t+\mathbf{v}(\mathbf{r},t)\cdot\boldsymbol{\nabla})\mathbf{v}(\mathbf{r},t)=e[\mathbf{E}(\mathbf{r},t)+\mathbf{v}(\mathbf{r},t)\times\mathbf{B}(\mathbf{r},t)]-\boldsymbol{\nabla}V_\Psi(\mathbf{r},t)\label{Cleb7}
\end{eqnarray} which contains the Lorentz force term $e[\mathbf{E}(\mathbf{r},t)+\mathbf{v}(\mathbf{r},t)\times\mathbf{B}(\mathbf{r},t)]$ acting upon the particle where $\mathbf{E}(\mathbf{r},t)=-\partial_t \mathbf{A}(\mathbf{r},t)-\boldsymbol{\nabla}V(\mathbf{r},t)$  and $\mathbf{B}(\mathbf{r},t)=\boldsymbol{\nabla}\times\mathbf{A}(\mathbf{r},t)$ are the electric  and magnetic fields respectively.  
Note in particular that the vorticity of the velocity field is non-zero and linked to the magnetic field by $m\boldsymbol{\Omega}(\mathbf{r},t)=-e\mathbf{B}(\mathbf{r},t)$.  Returning to our Clebsch potentials, the electromagnetic analogy suggests the identification:
   \begin{eqnarray}
e\mathbf{A}_{eff.}(\mathbf{r},t)=-{\alpha}(\mathbf{r},t)\boldsymbol{\nabla}{\beta}(\mathbf{r},t)\nonumber\\
eV_{eff.}(\mathbf{r},t)={\alpha}(\mathbf{r},t)\partial_t{\beta}(\mathbf{r},t)\nonumber\\
e\mathbf{E}_{eff.}(\mathbf{r},t)=\partial_t\alpha(\mathbf{r},t)\boldsymbol{\nabla}{\beta}(\mathbf{r},t)-\partial_t\beta(\mathbf{r},t)\boldsymbol{\nabla}{\alpha}(\mathbf{r},t)\nonumber\\
e\mathbf{B}_{eff.}(\mathbf{r},t)=-\boldsymbol{\nabla}{\alpha}(\mathbf{r},t)\times\boldsymbol{\nabla}{\beta}(\mathbf{r},t)
\label{Cleb8}
\end{eqnarray}
where we have introduced effective potentials associated with $\alpha$ and $\beta$ (note that this constitutes a quadrivector $eA_{eff.}^\mu(x)=\alpha(x)\partial^\mu\beta(x)$ where $A^\mu:=[V,\mathbf{A}]$, $\mathbf{x}:=[t,\mathbf{x}]$, $\partial^\mu=[\partial_t, -\boldsymbol{\nabla}]$). This leads immediately to the generalized Madelung equations for the fluid with vorticity in the presence of effective electromagnetic fields and real fields $\mathbf{A}, V$: 
 \begin{eqnarray}\left\{\begin{array}{ll}
-\partial_tS(\mathbf{r},t)=\frac{[\boldsymbol{\nabla}S(\mathbf{r},t)+{\alpha}(\mathbf{r},t)\boldsymbol{\nabla}{\beta}(\mathbf{r},t)-e\mathbf{A}(\mathbf{r},t)]^2}{2m}+V_\Psi(\mathbf{r},t)+{\alpha}(\mathbf{r},t)\partial_t{\beta}(\mathbf{r},t)+eV(\mathbf{r},t)\\
\partial_t R^2(\mathbf{r},t)+\boldsymbol{\nabla}\cdot(R^2(\mathbf{r},t)\mathbf{v}(\mathbf{r},t))=0\\
m\mathbf{v}(\mathbf{r},t)=\boldsymbol{\nabla}S(\mathbf{r},t)+{\alpha}(\mathbf{r},t)\boldsymbol{\nabla}{\beta}(\mathbf{r},t)-e\mathbf{A}(\mathbf{r},t)
\end{array}\right.
\label{Cleb9}
 \end{eqnarray}
These equations are completed by the conditions \ref{Cleb4} and we deduce the set of fundamental equations proposed by Sch\"onberg:
 \begin{eqnarray}\left\{\begin{array}{ll}
i\partial_t\Psi(\mathbf{r},t)=\frac{-1}{2m}[\boldsymbol{\nabla}+i{\alpha}(\mathbf{r},t)\boldsymbol{\nabla}{\beta}(\mathbf{r},t)-ie\mathbf{A}(\mathbf{r},t)]^2\Psi(\mathbf{r},t)+[{\alpha}(\mathbf{r},t)\partial_t{\beta}(\mathbf{r},t)+eV(\mathbf{r},t)]\Psi(\mathbf{r},t)\\
(\partial_t+\mathbf{v}(\mathbf{r},t)\cdot\boldsymbol{\nabla})\alpha(\mathbf{r},t):=\frac{d}{dt}\alpha(\mathbf{r}(t),t)=0\\
(\partial_t+\mathbf{v}(\mathbf{r},t)\cdot\bm{\nabla})\beta(\mathbf{r},t):=\frac{d}{dt}\beta(\mathbf{r}(t),t)=0\label{Cleb9b}
\end{array}\right.
 \end{eqnarray}
 Explained in another way: The Sch\"onberg theory we have just described allows us to find solutions to the pair of hydrodynamic equations
\begin{eqnarray}
\left\{\begin{array}{ll}
m(\partial_t+\mathbf{v}(\mathbf{r},t)\cdot\boldsymbol{\nabla})\mathbf{v}(\mathbf{r},t)=e[\mathbf{E}(\mathbf{r},t)+\mathbf{v}(\mathbf{r},t)\times\mathbf{B}(\mathbf{r},t)]-\boldsymbol{\nabla}V_\Psi(\mathbf{r},t)\\
\partial_t R^2(\mathbf{r},t)+\boldsymbol{\nabla}\cdot(R^2(\mathbf{r},t)\mathbf{v}(\mathbf{r},t))=0
\end{array}\right.
 \end{eqnarray}
 in the presence of vorticity $\boldsymbol{\Omega}\neq 0$. The condition for this is the introduction of two Clebsch fields $\alpha,\beta$ nonlinearly coupled to the $\Psi$ wave function according to Eq.~\ref{Cleb4} and with the generalized guidance condition $m\mathbf{v}(\mathbf{r},t)=\boldsymbol{\nabla}S(\mathbf{r},t)+{\alpha}(\mathbf{r},t)\boldsymbol{\nabla}{\beta}(\mathbf{r},t)-e\mathbf{A}(\mathbf{r},t)$.
 Note that Eq.~\ref{Cleb9b} or equivalently Eqs.~\ref{Cleb4},\ref{Cleb9} can be derived using a variational principle based on the Lagrangian 
 \begin{eqnarray}
 \mathcal{L}=-R^2[\partial_tS+{\alpha}\partial_t{\beta}+eV+\frac{[\boldsymbol{\nabla}S+{\alpha}\boldsymbol{\nabla}{\beta}-e\mathbf{A}]^2}{2m}]-\frac{(\boldsymbol{\nabla}R)^2}{2m}\label{Lagra}
 \end{eqnarray}
Several important properties follow from this dynamics. First note that we have $\mathbf{E}_{eff.}(\mathbf{r},t)+\mathbf{v}(\mathbf{r},t)\times\mathbf{B}_{eff.}(\mathbf{r},t)=0$ from which we deduce $\mathbf{E}_{eff.}(\mathbf{r},t)\cdot\mathbf{B}_{eff.}(\mathbf{r},t)=0$ and the effective Lorentz force $\mathbf{F}_{eff}=e[\mathbf{E}_{eff.}(\mathbf{r},t)+\mathbf{v}(\mathbf{r},t)\times\mathbf{B}_{eff.}(\mathbf{r},t)]$ therefore vanishes. Newton's Bohm dynamical law reads thus 
\begin{eqnarray}
m(\partial_t+\mathbf{v}(\mathbf{r},t)\cdot\boldsymbol{\nabla})\mathbf{v}(\mathbf{r},t)=e[\mathbf{E}(\mathbf{r},t)+\mathbf{v}(\mathbf{r},t)\times\mathbf{B}(\mathbf{r},t)]-\boldsymbol{\nabla}V_\Psi(\mathbf{r},t)\label{Cleb7b}
\end{eqnarray} as in Eq.~\ref{Cleb7b}.  A second remark concerns the vorticity of the velocity field which implies:
\begin{eqnarray}
m\boldsymbol{\Omega}(\mathbf{r},t)+e\mathbf{B}(\mathbf{r},t)=-e\mathbf{B}_{eff.}(\mathbf{r},t)=\boldsymbol{\nabla}{\alpha}(\mathbf{r},t)\times\boldsymbol{\nabla}{\beta}(\mathbf{r},t)\label{Cleb2b}
\end{eqnarray} 
Moreover, from Eq.~\ref{Cleb8} it is immediately clear that we have the first set of Maxwell's equations:
\begin{eqnarray}
\partial_t\mathbf{B}_{eff.}(\mathbf{r},t)=-\boldsymbol{\nabla}\times\mathbf{E}_{eff.}(\mathbf{r},t)\\
\boldsymbol{\nabla}\cdot\mathbf{B}_{eff.}(\mathbf{r},t)=0
\end{eqnarray}
which implies the formula $\partial_t\mathbf{B}_{eff.}(\mathbf{r},t)=\boldsymbol{\nabla}\times[\mathbf{v}(\mathbf{r},t)\times\mathbf{B}_{eff.}(\mathbf{r},t)]$. Therefore,  we obtain the hydrodynamic equation for the vorticity field $\bm{\Omega}$:
\begin{eqnarray}
\partial_t[m\boldsymbol{\Omega}(\mathbf{r},t)+e\mathbf{B}(\mathbf{r},t)]=\boldsymbol{\nabla}\times\left(\mathbf{v}(\mathbf{r},t)\times[m\boldsymbol{\Omega}(\mathbf{r},t)+e\mathbf{B}(\mathbf{r},t)]\right)
\end{eqnarray} which generalizes Eq.~\ref{12}.\\
To conclude this section, it's important to note that the Clebsch parameters introduced here don't necessarily form a global representation of the field in the entire configuration space (here 3-dimensional). In fact, Pfaff Darboux's theorem is only valid locally \cite{Scholle,Stern,Clebsch,Hankel}. In some cases, it may be necessary to introduce more than one pair of Clebsch potentials (at least $3N-1$ pairs of parameters are needed in a $3N$-dimensional configuration space \cite{Yoshida}). This is linked to the topology and helicity of the velocity field, which we won't discuss here (see \cite{Scholle,Stern,Moffatt}).   
%%%%%%%%%%%%%
\subsection{The quantum Rankine vortex}
\begin{figure}[h]
\includegraphics[width=12cm]{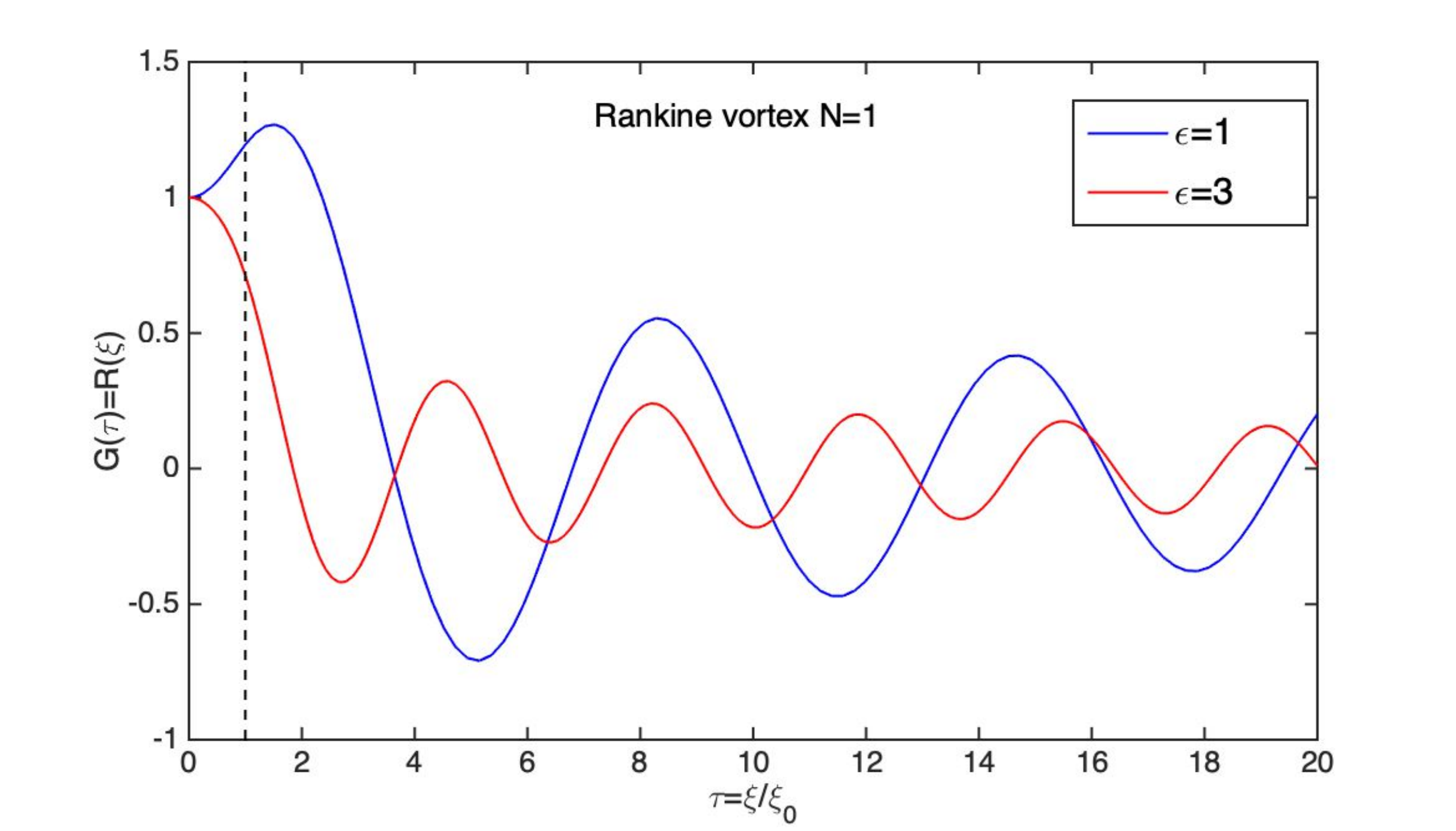} 
\caption{Radial profile $R(\xi)=G(\tau)$ of the wave function solution of Eq. \ref{eqrad4} for the choice $N=1$ and for two values of the normalized energy $\varepsilon$. The dotted vertical line separates the inner ($\tau<1$) and outer ($\tau>1$) domains of the quantum Rankine vortex. Vorticity is constant in the inner domain, and cancels out in the outer domain. } \label{Figure1}
\end{figure}

To illustrate Sch\"onberg's theory, let's consider the case of a vorticity tube, which is the quantum analogue of Rankine's classical hydrodynamic model for an idealized tornado.  In this model we suppose no real electromagnetic field and the vorticity is induced by the presence of effective fields according to Eq.~\ref{Cleb8}.   For a vortex with cylindrical  symmetry around the $z$ axis we use the Clebsch potentials
\begin{eqnarray}
\alpha(\mathbf{r},t)=f(\xi)\nonumber\\
\beta(\mathbf{r},t)=\varphi+ g(t)
\end{eqnarray}  where $\xi=\sqrt{(x^2+y^2)},z,\varphi$ are cylindrical coordinates and $f(\xi)$, $g(t)$ are two unknown functions. This choice implies the effective fields:
    \begin{eqnarray}
e\mathbf{A}_{eff.}(\mathbf{r},t)=-{\alpha}(\mathbf{r},t)\boldsymbol{\nabla}{\beta}(\mathbf{r},t)=-\frac{f(\xi)}{\xi}\hat{\bm{\varphi}}\nonumber\\
eV_{eff.}(\mathbf{r},t)={\alpha}(\mathbf{r},t)\partial_t{\beta}(\mathbf{r},t)=f(\xi)\frac{dg(t)}{dt}\nonumber\\
e\mathbf{E}_{eff.}(\mathbf{r},t)=\partial_t\alpha(\mathbf{r},t)\boldsymbol{\nabla}{\beta}(\mathbf{r},t)-\partial_t\beta(\mathbf{r},t)\boldsymbol{\nabla}{\alpha}(\mathbf{r},t)=-\frac{df(\xi)}{d\xi}\frac{dg(t)}{dt}\hat{\bm{\xi}}\nonumber\\
e\mathbf{B}_{eff.}(\mathbf{r},t)=-\boldsymbol{\nabla}{\alpha}(\mathbf{r},t)\times\boldsymbol{\nabla}{\beta}(\mathbf{r},t)=-\frac{1}{\xi}\frac{df(\xi)}{d\xi}\hat{\mathbf{z}}
\label{Cleb8c}
\end{eqnarray}
This induces a vorticity  
\begin{eqnarray}
m\boldsymbol{\Omega}(\mathbf{r},t)=-e\mathbf{B}_{eff.}(\mathbf{r},t)=\frac{1}{\xi}\frac{df(\xi)}{d\xi}\hat{\mathbf{z}}\label{Cleb2c}
\end{eqnarray} corresponding to a tornado surrounding the $z$ axis.\\
\indent To be consistent, we look for a velocity field given by the guidance formula $m\mathbf{v}(\mathbf{r},t)=\boldsymbol{\nabla}S(\mathbf{r},t)+{\alpha}(\mathbf{r},t)\boldsymbol{\nabla}{\beta}(\mathbf{r},t)$ such that the motion is azimuthal.  The wave function solution of Eq.~\ref{Cleb9b} will be assumed to be $\Psi(\mathbf{r},t):=R(\xi)e ^{iN\varphi} e^{-iEt}$ where $N\in \mathbb{R}$ describes the orbital momentum and $E$ the particle energy (note that $N$ is not necessarily an integer but this will be assumed later). The velocity field $\mathbf{v}=v_\varphi\hat{\bm{\varphi}}$ is then written 
\begin{eqnarray}
mv_\varphi(\xi)=\frac{N}{\xi}-eA_{eff,\varphi}(\xi)=\frac{N+f(\xi)}{\xi}\label{truc}
\end{eqnarray}
which can also be justified by calculating the velocity circulation on an integration contour around the $z$ axis. 
The energy conservation obtained from the generalized HJ equation \ref{Cleb9} reads $E=\frac{(mv_\varphi)^2}{2m}+eV_{eff}+V_\Psi$ and leads to the radial Schr\"odinger equation:
\begin{eqnarray}
0=2m[E-\frac{(N+f(\xi))^2}{2m\xi^2}-f(\xi)\frac{dg(t)}{dt}]R(\xi)+ \frac{1}{\xi}\frac{dR(\xi)}{d\xi}+\frac{d^2R(\xi)}{d\xi^2}\label{eqrad}
\end{eqnarray} Note that fluid  conservation is automatically preserved since $\bm{\nabla}\cdot[|\Psi|^2\mathbf{v}]=\frac{1}{\xi}\partial_\varphi[v_\varphi(\rho)R^2(\xi)]=0$. 
To further constrain the model we need Eq.~\ref{Cleb4}.  The condition $\frac{d}{dt}\alpha(\mathbf{r}(t),t)=0$ is automatically fulfilled but the constraint $\frac{d}{dt}\beta(\mathbf{r}(t),t)=0$ reads:
\begin{eqnarray}
0=\frac{dg(t)}{dt}+\frac{v_\varphi(\xi)}{\xi}
\end{eqnarray} 
After combining with Eq.~\ref{truc} we obtain 
\begin{eqnarray}
\frac{dg(t)}{dt}=-\frac{v_\varphi(\xi)}{\xi}=-\frac{N+f(\xi)}{m\xi^2}\label{cons}
\end{eqnarray}
Moreover, in order Eq.~\ref{cons} to be true  $N+f(\xi)$ must be proportional to $\xi^2$. This is only possible for a constant effective magnetic field $-eB_0=\frac{1}{\xi}\frac{df(\xi)}{d\xi}$ which after integration leads to
\begin{eqnarray}
f(\xi)=\frac{-eB_0\xi^2}{2}+f_0= \frac{-eB_0\xi^2}{2}-N
\end{eqnarray} where the constant $f_0$ must be $-N$ to satisfy Eq.~\ref{cons}.  We deduce $g(t)=\frac{eB_0}{2m}t=-\tilde{\omega} t$  and therefore $\beta=\varphi+\tilde{\omega} t$ where $\tilde{\omega}$ is an angular Larmor frequency for the Clebsch potential $\beta$. We have 
\begin{eqnarray}
v_\varphi(\xi)=-\frac{eB_0}{2m}\xi=\tilde{\omega}\xi\nonumber\\
\Omega=-\frac{eB_0}{m}=2\tilde{\omega}\label{inside}
\end{eqnarray} The constant vorticity $\Omega$ is twice the angular frequency $\tilde{\omega}$ a property already obtained with the classical Rankine tornado.
In the end this allows us to rewrite Eq.~\ref{eqrad} as:
 \begin{eqnarray}
0=[2mE+NeB_0+\frac{e^2B_0^2\xi^2}{4}]R(\xi)+ \frac{1}{\xi}\frac{dR(\xi)}{d\xi}+\frac{d^2R(\xi)}{d\xi^2}\label{eqradbis}
\end{eqnarray}
\indent To complete the description of our tornado, we assume that the vorticity is constant and confined within a tube of radius $\xi_0$ and such that $\Omega=0$ outside the tube (or filament). In the external domain, the fluid is irrotational, so let's assume $\alpha=\beta=0$ for $\xi>\xi_0$.  According to the guiding formula, the velocity field in this domain is given by the standard de Broglie-Bohm formula $m\mathbf{v}(\mathbf{r},t)=\boldsymbol{\nabla}S(\mathbf{r},t)$ with the wave function $\Psi'(\mathbf{r},t):=R'(\xi)e ^{iN'\varphi} e^{-iE't}$ as above, with $N'$ and $E'$ apriori different from $N$ and $E$ in the inner domain ($\xi<\xi_0$). Note that $N' \in \mathbb{Z}$ as usually assumed.
We thus have:
\begin{eqnarray}
mv_\varphi(\xi)=\frac{N'}{\xi}\label{trucBis}
\end{eqnarray}
From the energy conservation (i.e., HJ equation) we have 
$E'=\frac{(mv_\varphi)^2}{2m}+V_\Psi$ and deduce the radial Schr\"odinger equation:
\begin{eqnarray}
0=2m[E'-\frac{N^2}{2m\xi^2}]R'(\xi)+ \frac{1}{\xi}\frac{dR'(\xi)}{d\xi}+\frac{d^2R'(\xi)}{d\xi^2}\label{eqradBis}
\end{eqnarray}
The two domains of the vortex are connected by assuming the continuity of the velocity field at $\xi=\xi_0$ which imposes by comparing  \ref{inside} and \ref{trucBis} the equality $v_\varphi(\xi_0)=\frac{N'}{m\xi_0}=-\frac{eB_0}{2m}\xi_0$ and therefore the constraint
\begin{eqnarray}
N'=-\frac{eB_0}{2}\xi_0^2
\label{tructri}
\end{eqnarray}
Moreover, by imposing the continuity of $E$ and $N$, which is equivalent to assuming the continuity of the phase of the wave function, we have $N=N'$ and $E=E'$. The two radial equations \ref{eqradbis} and \ref{eqradBis} are greatly simplified, since the potential energy terms and their radial first derivative are also continuous. We can thus group \ref{eqradbis} and \ref{eqradBis} into 
\begin{eqnarray}
0=[2mE+N^2\frac{(\xi^2-2\xi_0^2)}{\xi_0^4}\Theta(\xi_0-\xi)-\frac{N^2}{\xi^2}\Theta(\xi-\xi_0)]R(\xi)+ \frac{1}{\xi}\frac{dR(\xi)}{d\xi}+\frac{d^2R(\xi)}{d\xi^2}\label{eqradtri}
\end{eqnarray} where we removed the notation $R'(\xi)$ for the external domain since we have a single continuous radial wave function for the whole space [$\theta(x)$ is the Heaviside function with $\theta(x)=1$ for $x\geq 0$ and zero otherwise]. To solve this equation it is convenient to use the variables $\tau:=\xi/\xi_0$ and $R(\xi):=G(\tau)$ and Eq.~\ref{eqradtri} reads: 
\begin{eqnarray}
0=[\varepsilon+N^2(\tau^2-2)\Theta(1-\tau)-\frac{N^2}{\tau^2}\Theta(\tau-1)]G(\tau)+ \frac{1}{\tau}\frac{dG(\tau)}{d\tau}+\frac{d^2G(\tau)}{d\tau^2}\label{eqrad4}
\end{eqnarray} with $\varepsilon=2mE\xi_0^2$. In the domain $\tau>1$, outside the vortex, \ref{eqrad4} reduces to Bessel equation with  the general solution \begin{eqnarray}
R(\xi)=G(\tau)=C_1 J_N(\sqrt{\varepsilon }\tau)+C_2 Y_N(\sqrt{\varepsilon} \tau)=C_1 J_N(\sqrt{2mE}\xi)+C_2 Y_N(\sqrt{2mE}\xi)
\end{eqnarray} where $J_N$ and $Y_N$ are the Bessel functions of the first and second kind respectively and $C_1,C_2$ two constants.  Note that in absence of vorticity core, i.e., if $B_0=0$, the only physical solution is the usual Bessel function $J_N(\sqrt{2mE}\xi)$.    In particular, if $N>0$ we have $J_N(0)=0$ and the wave function indeed vanishes along the $z$ axis in order to agree with the existence of the phase singularity on the nodal line. If we now assume $B_0\neq 0$ the solution $G(\tau)$ near the origin is given by the expansion \begin{eqnarray}
G(\tau)=C_0(1+\frac{2N^2-\varepsilon}{4}\tau^2+O(\tau^3))\label{orig}
\end{eqnarray} with $C_0$ a constant. Note that the sign of the curvature at the origin is negative  only if $\varepsilon>2N^2$, i.e., $E>\frac{N^2}{m\xi_0^2}$. This is clearly understood if we write $G(\tau)=W(\tau)\sqrt{\tau}$ leading  to the Schr\"odinger equation    
\begin{eqnarray}
0=[\varepsilon+N^2(\tau^2-2)\Theta(1-\tau)-\frac{N^2-\frac{1}{4}}{\tau^2}\Theta(\tau-1)]W(\tau)+\frac{d^2W(\tau)}{d\tau^2}\label{eqrad5}
\end{eqnarray} This corresponds to the problem of a particle interacting with a one-dimensional potential barrier $U_{eff}(\tau)=N^2(2-\tau^2)\Theta(1-\tau)+\frac{N^2-\frac{1}{4}}{\tau^2}\Theta(\tau-1)$ whose value decreases monotonically towards zero for $\tau\rightarrow +\infty$ and such that the maximum value  $U_{eff}(0)=2N^2$ is reached at the origin $\tau=0$. From this, we deduce that Eq. \ref{eqrad5}, and therefore \ref{eqrad4}, admits a convergent solution only if $\varepsilon>0$. On the other hand, if $\varepsilon > U_{eff}(0)=2N^2$ then the energy is greater than the barrier and the wave function must decrease in amplitude. In the opposite case, if $0<\varepsilon<2N^2$, the wave function must enter the barrier by tunneling, which explains qualitatively the behavior at origin in Eq. \ref{orig}.\\
\indent A numerical solution to Eq. \ref{eqrad4} is obtained by setting $G(0)=C_0=1$, $\frac{d}{d\tau}G(0)=0$ and using a Runge-Kutta 4 routine on Matlab. Fig. \ref{Figure1}  shows two solutions for the $N=1$ case and corresponding to $\varepsilon=3>2$ and $\varepsilon=1 <2$ respectively.   
%%%%%%%%%%

\begin{figure}[h]
\includegraphics[width=10cm]{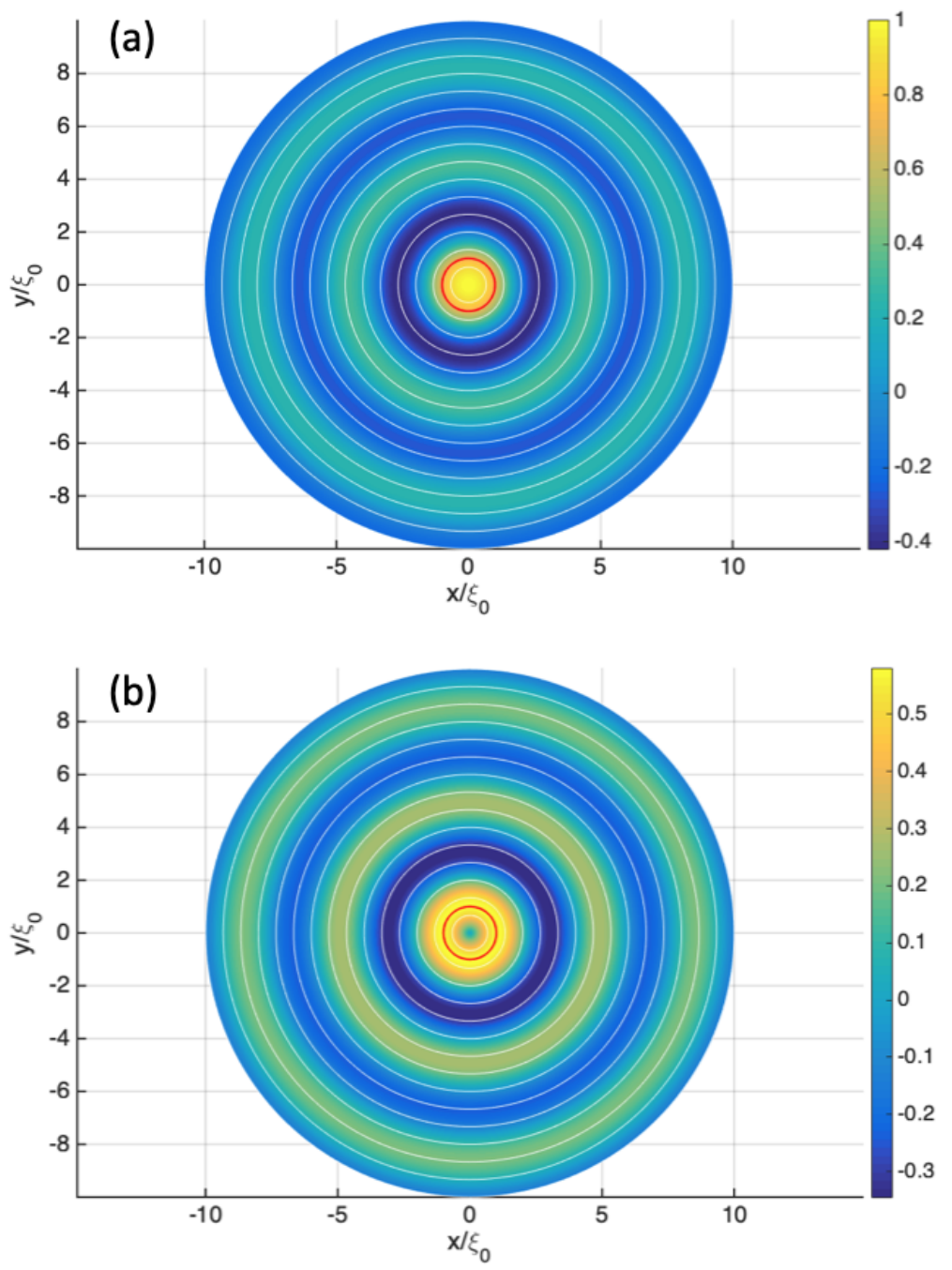} 
\caption{(a) Two-dimensional profile of the quantum Rankine vortex (for $N=1$) in the $x-y$ plane compared with (b) the profile of the ideal Bessel vortex (corresponding to $B_0=0$ and zero vorticity).  The concentric circles are de Broglie Bohm trajectories. The red circle corresponds to the critical radius $\xi=\sqrt{x^2+y^2}=\xi_0$ separating the inner and outer domains of the Rankine vortex. } \label{Figure2}
\end{figure}

%%%%%%%
The velocity field is therefore cylindrically symmetrical, surrounding the core of the vortex.  The resulting quantum system is virtually indistinguishable from the irrotational case, in which the vortex is a nodal line (specifically if $\xi_0\rightarrow 0$). Comparison of the quantum fields is shown in Figure \ref{Figure2}.   
%%%%%%%%%%

%%%%%%%
It's important to note that the discussion here of the Rankine vortex (in its quantum version) is in line with the results obtained in pure classical hydrodynamics for a Eulerian fluid.  Moreover, in the usual derivation of Rankine vortex properties, the pressure $p(\mathbf{r},t)$ can be obtained by integration of the Newton-Euler law. The same is possible here for the quantum potential $V_\psi(\mathbf{r},)$ replacing the pressure field: Starting from the formula $m\frac{v_\varphi^2}{\xi}=\frac{\partial}{\partial \xi}V_\Psi(\xi)$ and using Eq. \ref{inside} and \ref{trucBis} we can by integration obtain the quantum potential in the two domains $\xi<\xi_0$ and $\xi>\xi_0$. By assuming continuity of these potentials at the core-vortex boundary $\xi=\xi_0$ we get:
\begin{eqnarray}
V_\Psi(\xi)=\frac{N^2}{2m\xi_0^4}(\xi^2-2\xi_0^2)\Theta(\xi_0-\xi)-\frac{N^2}{2m\xi^2}\Theta(\xi-\xi_0)+C
\end{eqnarray} where $C$ is a constant.  The same result is directly obtained from our general HJ equation \ref{eqradtri} which fixes the constant $C=E$.
\subsection{Remarks}
\indent To conclude this analysis of Sch\"onberg's theory, several remarks are  useful and important. First of all, it's important to remember that the primary motivation for this theory is to extend Schr\"odinger's quantum theory, based on Madelung de Broglie's hydrodynamic analogy.   Based on the idea that the description of a quantum fluid endowed with $\Omega\neq$ vorticity makes sense, we are directly led through the introduction of Clebsch potentials whose coupling with the $\Psi$ wavefunction is non-linear and non-trivial. This is particularly apparent in the Lagrangian function \ref{Lagra}. Takabayasi suggested that the introduction of these Clebsch potentials might have something to do with the theory of nuclear forces and mesons, which in the 1950s had reached an impasse. However for us, what's important is that in the de Broglie Bohm theory, Clebsch potentials extend the class of possible particle motions beyond the standard HJ guiding formula limited to irrotational motions.\\
\indent It's interesting in this context to recall that Dirac reached similar conclusions in his classical theory of the electron \cite{Dirac}. Indeed, limiting ourselves to standard HJ theory based on irrotational motion, we must have the guiding formula $m\mathbf{v}(\mathbf{r},t)=\boldsymbol{\nabla}S(\mathbf{r},t)-e\mathbf{A}(\mathbf{r},t)$ in the presence of a magnetic field.  It is clearly impossible within the framework of this classical theory to imagine a probabilistic cloud of electrons which at a given instant $t$ would be motionless in a magnetic field (because then from $\bm{\Omega}=0$ we'd have $B=0$).  This shows that HJ theory is limited within the classical framework and is only a subclass of Newtonian theory. The hydrodynamic formalism makes all this clear.   The irrotationality postulate $\Omega=0$ is not general enough.   So it seems natural to assume that the same applies to de Broglie Bohm's quantum theory.  If, as we believe, following de Broglie, this theory is the natural completion of classical theory (due to its ontological clarity and historical continuity with the work of Hamilton, Jacobi and others), then the introduction of Clebsch potentials seems self-evident.\\
\indent From this point of view, the de Broglie-Bohm theory is not only an ontological approach that gives meaning to quantum mechanics; it is also a means of extending or anticipating possible extensions of quantum mechanics. In other words, it becomes useful for imagining new forms of physics beyond current quantum theory.    Of course, one question we must ask ourselves here is why we don't see the presence of these Clebsch $\alpha$ and $\beta$ potentials in the laboratory? Two suggestions are possible here. Firstly, in line with the literature on hydrodynamics and the Kelvin-Helmholtz theorem, we can assume that quantum fluids are currently strongly dominated by the irrotational regime $\Omega=0$. In fact, in Eulerian fluid mechanics, vortices are particular regions of space where this vorticity is confined to the core region. By analogy with this work, we can assume that the same applies to Bohmian theory. The example of the Rankine vortex shows that indeed this region of non-zero vorticity can be completely isolated from the environment, where the quantum fluid can be considered as irrotational.\\
\indent It's also important to note that vortices associated with nodal lines in the usual Schr\"odinger theory (without Clebsch potentials) are very difficult regions to study and probe experimentally, since in these regions the probability of presence given by Born's rule tends towards zero.  This is where the link with the work of Valentini and Bohm-Vigier comes in. It's entirely possible that regions of non-zero vorticity were created in the early universe (close to the Big Bang). These regions could still have a detectable effect via their effect on particle dynamics. Moreover, vortex nodal lines obeying the standard Schr\"odinger equation (i.e. in the absence of $\alpha$ and $\beta$) are known to be highly chaotic regions for Bohmian dynamics. In Valentini theory, where relaxation to quantum equilibrium is paramount, the presence of vortices plays an essential role in monitoring relaxation to quantum equilibrium and Born's law.  However, the presence of non-zero vorticity domains (sort of Bohmian cosmic strings) could disturb these processes and could represent regions of spacetime where particles out of quantum equilibrium are captured and trapped.  This deserves further investigation.\\
\indent  Another point that deserves a more extensive analysis (but which we will only touch on briefly because of its great complexity) concerns the link between Clebsch's formalism and Pauli's theory of $1/2$ spin particles developed by Bohm Schiller and Tiomno in a Hydrodynamic form \cite{Tiomno,Holland,BohmHiley}. In Pauli's theory, single electrons are described by $\Psi(\mathbf{r},t)=\left(\begin{array}{ll}\Psi_{\uparrow}(\mathbf{r},t)\\ \Psi_{\downarrow}(\mathbf{r},t)
\end{array}\right)$ two-component spinors, of which we can give a hydrodynamic representation. To do this, we write 
\begin{eqnarray}
\Psi(\mathbf{r},t)=\left(\begin{array}{ll}\Psi_{\uparrow}(\mathbf{r},t)\\ \Psi_{\downarrow}(\mathbf{r},t)
\end{array}\right)=R(\mathbf{r},t)e^{iS(\mathbf{r},t)}\left(\begin{array}{ll}\cos{(\frac{\vartheta(\mathbf{r},t)}{2})}e^{-i\frac{\varphi(\mathbf{r},t)}{2}(\mathbf{r},t)}\\ \sin{(\frac{\vartheta(\mathbf{r},t)}{2})}e^{+i\frac{\varphi(\mathbf{r},t)}{2}(\mathbf{r},t)}
\end{array}\right)
\end{eqnarray}
 which contains 4 real fields $R$, $S$, $\vartheta$, $\vartheta$ with a clear kinetic and dynamic interpretation. The simplest quantity is $R$ defined by $\Psi^\dagger(\mathbf{r},t)\Psi(\mathbf{r},t)=R^2(\mathbf{r},t)$ and which is associated, like in the usual de Broglie bohm theory, with the Born probability distribution (i.e., assuming quantum equilibrium). The two internal angles $\vartheta,\varphi$ define the local orentation of the spinor field. In fact, local spin can be introduced by the formula 
 \begin{eqnarray}
 \bm{\Sigma}(\mathbf{r},t)=\frac{1}{2}\frac{\Psi^\dagger(\mathbf{r},t)\bm{\sigma}\Psi(\mathbf{r},t)}{\Psi^\dagger(\mathbf{r},t)\Psi(\mathbf{r},t)}=\frac{1}{2}\hat{\mathbf{n}}(\mathbf{r},t)
 \end{eqnarray}
 which involves the Pauli matrices $\sigma_x,\sigma_y, \sigma_z$ grouped together in the form of a vector operator $\bm{\sigma}=\sigma_x\hat{\mathbf{x}}+\sigma_y\hat{\mathbf{y}}+\sigma_z\hat{\mathbf{z}}$ and  whose direction $\hat{\mathbf{n}}=\left[\begin{array}{ll}\cos{\varphi}\sin{\vartheta}\\ \sin{\varphi}\sin{\vartheta}\\\cos{\vartheta}\end{array}\right]$ (in spherical coordinates) is a unit vector characterized by the internal angles $\vartheta,\varphi$ which vary continuously in the space-time of the Pauli field. The last parameter, the phase $S(\mathbf{x},t)$, is clearly a generalization of the HJ action and, in particular, it can be shown that the Bohmian probability fluid velocity, which also defines the particle velocity, is given by 
  \begin{eqnarray}
 \mathbf{v}(\mathbf{r},t)=\bm{\nabla}\frac{S(\mathbf{r},t)}{m}-\frac{e}{m}\mathbf{A}(\mathbf{r},t)-\frac{1}{2m}\cos{\vartheta}\bm{\nabla}\varphi(\mathbf{r},t)+\frac{\bm{\nabla}\times\bm{\Sigma}(\mathbf{r},t)}{m\Psi^\dagger(\mathbf{r},t)\Psi(\mathbf{r},t)}\label{paulicurrent}
 \end{eqnarray}
In this expression, the first term and second term is the usual de Broglie Madelung velocityin presence of magnetic potential $\mathbf{A}(\mathbf{r},t)$. The fourth term is a magnetic term associated with the particle's spin current. It was omitted by Bohm Schiller and Tiomno  \cite{Tiomno} but must appear if we consider Pauli's theory as the non-relativistic limit of Dirac's equation \cite{BohmHiley}.  Finally, the third term is a special case of the Clebsch representation for the spin fluid with $\alpha=-\cos{\vartheta}$ and $\beta=\varphi/2$. All this suggests a link between the particle's spin and the Clebsch parameters, but this is neither certain nor obligatory, as we can also introduce these parameters without any reference to spin, for example in the Klein Gordon equation, which is associated with a zero-spin particle. Furthermore, as Sch\"onberg and Takabayasi showed \cite{Schonberg,Takabayasi}, the Clebsch potential formalism can be extended to the relativistic Dirac equation for $1/2$ spin particles, meaning that the  representation \ref{paulicurrent} can in fact be generalized by adding new Clebsch terms $\alpha\bm{\nabla}\beta$ not necessarily connected to spin. \\
\indent To conclude this section, following de Broglie, we started from the idea that classical HJ theory is transformed, in quantum physics, into de Broglie Bohm theory. Since HJ theory represents only one possible sub-class of Newtonian mechanical motion (i.e. the class of irrotational motions with zero vorticity), it becomes natural to look in the quantum domain for the broadest completion corresponding to motions not necessarily limited by the zero vorticity constraint.  This is the theory proposed by Sch\"onberg and Takabayasi \cite{Schonberg,Takabayasi}, and offers a natural extension of the Schr\"odinger equation. Such a completion would clearly be impossible to formulate in the minimalist description of Bell and DGZ, which is limited to a formalism without quantum potential.  This shows once again the importance of mechanical analogies and historical links with methods developed in the 19$^{th}$ century in optics, mechanics and hydrodynamics.

%%%%%%%%%%%
\section{Conclusion}\label{section7}
To conclude this review. We have tried to show in this work that the de Broglie Bohm theory, whose centenary we are celebrating, is extremely rich in physics and mathematics. This theory is strongly based on analogies between optics, fluid mechanics and the quantum theory intuited by Louis de Broglie as early as 1923-25.  Clearly, we don't believe that this theory can be reduced to the minimal form taught and popularized in recent years on the basis of Bell's work (even if this minimalist approach may be of pedagogical interest in certain cases).   In fact, de Broglie Bohm's theory draws heavily on the work of Hamilton and Jacobi, and even earlier on the work of Maupertuis and Lagrange in mechanics and Fermat in optics. All this suggests the importance of hydrodynamic description in understanding and, if possible, extending de Broglie Bohm's theory.   We have considered different scenarios for the possible extension of de Broglie Bohm theory, either at the statistical or dynamic level. All are interesting and open up possibilities for a better understanding of quantum mechanics. Basically, the underlying idea is that defended by Bell when he told: 
\begin{quote}
`\textit{I'm quite convinced of that: quantum theory is only a temporary expedient}' \cite{Bellinterview}. 
\end{quote}

\indent An important element of this review on the different ways or alternatives of completing or modifying the de Broglie Bohm theory is that they are not really independent. For example, the idea of introducing dynamical fluctuations into the laws of motion could affect the way quantum equilibrium is discussed. Similarly, the idea of introducing vorticity into the quantum fluid could affect the way we see particles (according to Bohm and Vigier \cite{Vigier}, for example, we could envisage, in agreement with Takabayasi and Sch\"onberg \cite{Takabayasi,Schonberg}, that particles are kind of mini vortex of very small dimensions carried by the fluid forming a sub-quantum medium and affected by Brownian fluctuations). Last but not least, this makes a deep connection with de Broglie's double solution approach \cite{debroglie1956,Drezet1}, as vortices and other topological structures within the fluid are good candidates for soliton models. In fact, what we see is that all these approaches are part of the same general project: that of better understanding quantum mechanics in order to go beyond it. 

%%%%%%%%%%%%%%%%%%%%%%%%%%%%%%%%%%%%%%%%%%

\end{document}